\documentclass[prd,showpacs,preprintnumbers,amsmath,amssymb]{revtex4}
\usepackage{graphics}
\usepackage{epsfig}
\usepackage{subfigure}
\usepackage{dcolumn}% Align table columns on decimal point
\usepackage{bm}% bold math
\usepackage{color}

\begin{document}
\title{Investigating possible decay modes of $Y(4260)$ under the $D_1(2420)\bar D +c.c$ molecular state ansatz}

\author{Gang Li$^{1}$\footnote{gli@mail.qfnu.edu.cn} and
Xiao-Hai Liu$^{2}$\footnote{liuxiaohai@pku.edu.cn}}

\affiliation{1) Department of Physics, Qufu Normal University, Qufu
273165, People's Republic of China}

\affiliation{2) Department of Physics and State Key Laboratory of
Nuclear Physics and Technology, Peking University, Beijing 100871,
People's Republic of China}

%\affiliation{3) Institute of High Energy Physics, Chinese Academy of
%Sciences, Beijing 100049, P.R. China}

%\affiliation{4) Theoretical Physics Center for Science Facilities,
%Chinese Academy of Sciences, Beijing 100049, P.R. China}

\begin{abstract}
By assuming that $Y(4260)$ is a $D_1\bar D$ molecular state, we
investigate some hidden-charm and charmed pair decay channels of
$Y(4260)$ via intermediate $D_1\bar D$ meson loops with an effective
Lagrangian approach. Through investigating the $\alpha$-dependence
of branching ratios and ratios between different decay channels, we
show that the intermediate $D_1 \bar D$ meson loops are crucial for
driving these transitions of $Y(4260)$ studied here. The coupled
channel effects turn out to be more important in $Y(4260) \to
D^{*}\bar{D}^{*}$, which can be tested in the future experiments.
\end{abstract}

\date{\today}


%14.40.Rt Exotic mesons

%13.75.Lb Meson-meson interactions

%13.20.Gd Decays of J/¦×, ¦´, and other quarkonia

%14.40.Pq Heavy quarkonia

%14.40.Lb Charmed mesons

\maketitle

\section{Introduction}
\label{sec:introduction}

During the past years, the experimental observation of a large
number of so-called $XYZ$ states has initiated tremendous efforts to
unravel their nature beyond the conventional quark model (for recent
reviews, see, e.g.
Refs~\cite{Swanson:2006st,Eichten:2007qx,Voloshin:2007dx,Godfrey:2008nc,Drenska:2010kg}).
$Y(4260)$ was reported by the BaBar Collaboration in the $\pi^+\pi^-
J/\psi$ invariant spectrum  in $e^+e^- \to \gamma_{ISR}\pi^+\pi^-
J/\psi$~\cite{Aubert:2005rm}, which has been confirmed both by the
CLEO and Belle collaboration~\cite{He:2006kg,Yuan:2007sj}. Its mass
and total width are well determined as $m=4263^{+8}_{-9}$ MeV and
$\Gamma_Y=95\pm 14$ MeV, respectively~\cite{Beringer:1900zz}. The
new datum from BESIII confirms the signal in $Y(4260)\to
J/\psi\pi^+\pi^-$ with much higher
statistics~\cite{Ablikim:2013mio}. The mass of $Y(4260)$ does not
agree to what is predicted by the potential quark model. Further
more, the most mysterious fact is that as a charmonium state with
$J^{PC}=1^{--}$, it is only ``seen" as a bump in the two pion
transitions to $J/\psi$, but not in any open charm decay channels
like $D \bar{D}$, $D^* \bar{D}+c.c.$ and $D^*\bar{D}^*$, or other
measured channels. The line shapes of the cross section for $e^+
e^-$ annihilations into $D^{(*)}$ meson pairs appear to have a dip
at its peak mass $4.26$ GeV instead of a bump.

Since the observation of the $Y(4260)$, many theoretical
investigations have been carried out (for a review see
Ref.~\cite{Brambilla:2010cs}). It has variously been identified as a
conventional $\psi(4S)$ based on a relativistic quark
model~\cite{LlanesEstrada:2005hz}, a tetraquark $c {\bar c} s {\bar
s}$ state~\cite{Maiani:2005pe}, a charmonium
hybrid~\cite{Zhu:2005hp,Kou:2005gt,Close:2005iz}, hadronic molecule
of
$D_1\bar{D}$~\cite{Ding:2007rg,Ding:2008gr,Wang:2013cya},${^{\footnotemark[1]}}$
~\footnotetext[1]{Notice that there are two $D_1$ states of similar
masses, and the one in question should be the narrower one, i.e. the
$D_1(2420)$ ($\Gamma=27$ MeV), the $D_1(2430)$($\Gamma \simeq 384$
MeV) is too broad to form a molecular
state~\cite{Filin:2010se,Guo:2011dd,Guo:2013zbw}.}
$\chi_{c1}\omega$~\cite{Yuan:2005dr},
$\chi_{c1}\rho$~\cite{Liu:2005ay}, $J/\psi
f_0$~\cite{MartinezTorres:2009xb}, a
cusp~\cite{vanBeveren:2009fb,vanBeveren:2009jk} or a non-resonance
explanation~\cite{vanBeveren:2010mg,Chen:2010nv} etc. The dynamical
calculation of tetraquark states indicated that $Y(4260)$ can not be
interpreted as P-wave $1^{--}$ state of charm-strange
diquark-antidiquark, because the corresponding mass is found to be
200 MeV heavier~\cite{Ebert:2005nc}. In Ref.~\cite{Zhang:2010mw},
the authors also studied the possibility of $Y(4260)$ as P-wave
$1^{--}$ state of charm-strange diquark-antidiquark state in the
framework of QCD sum rules and arrived the same conclusion as
Ref.~\cite{Ebert:2005nc}. Some lattice calculations give the mass of
vector hybrid within this mass region~\cite{Bali:2003tp}, which is
very close to the new charmonium-like state
Y(4360)~\cite{Wang:2007ea}. With the $D_1 \bar D$ molecular ansatz,
a consistent description of some of the experimental observations
can be obtained, such as its non observation in open charm decays,
or the observation of $Z_c(3900)$ as mentioned in
Ref.~\cite{Wang:2013cya}, the threshold behavior in its main decay
channels are investigated in Ref.~\cite{Liu:2013vfa} and the
production of $X(3872)$ is studied in the radiative decays of
$Y(4260)$~\cite{Guo:2013zbw}. Under such a molecular state
assumption, a consistent description of many experimental
observations could be obtained. However, as studied
in~\cite{Li:2013yka}, the production of an S-wave $D_1 {\bar D}$
pair in $e^+e^-$ annihilation is forbidden in the limit of exact
heavy quark spin symmetry, which substantially weakens the arguments
for considering the $Y(4260)$ charmonium-like resonance as a $D_1
{\bar D}$ molecular state.

The intermediate meson loop (IML) transition is one of the possible
nonperturbative dynamical mechanisms, especially when we investigate
the pertinent issues in the energy region of
charmonium~\cite{Li:1996yn,Cheng:2004ru,Anisovich:1995zu,Li:2011ssa,Li:2007ky,Wang:2012mf,Li:2012as,Li:2007xr,Achasov:1990gt,Achasov:1991qp,Achasov:1994vh,Achasov:2005qb,Zhao:2006dv,Zhao:2006cx,Wu:2007jh,Liu:2006dq,Zhao:2005ip,Li:2007au,Liu:2009dr,Zhang:2009kr,Liu:2009vv,Liu:2010um,Wang:2012wj,Guo:2009wr,Guo:2010zk,Guo:2010ak}.
During the last decade, many interesting observations were announced
by Belle, BaBar, CLEO, BESIII, and so on. And in theoretical study,
it is widely recognized that the IML may be closely related to a lot
of nonperturbative phenomena observed in
experiments~\cite{Achasov:1990gt,Achasov:1991qp,Achasov:1994vh,Achasov:2005qb,Wu:2007jh,Liu:2006dq,Cheng:2004ru,Anisovich:1995zu,Zhao:2005ip,Li:2007au,Liu:2009dr,Zhang:2009kr,Liu:2009vv,Liu:2010um,Wang:2012wj,Guo:2009wr,Guo:2010zk,Guo:2010ak,Li:2013xia,Brambilla:2004wf,Brambilla:2004jw},
e.g. apparent OZI-rule violations, sizeable non-$D\bar D$ decay
branching ratios for
$\psi(3770)$~\cite{Achasov:1990gt,Achasov:1991qp,Achasov:1994vh,Achasov:2005qb,Liu:2009dr,Zhang:2009kr},
the HSR violations in charmonium
decays~\cite{Liu:2009vv,Liu:2010um,Wang:2012wj}, the hidden
charmonium decays of the newly discovered $Z_c$~\cite{Li:2013xia},
etc.

In this work, we will investigate the hidden-charm decays of
$Y(4260)$  and $Y(4260) \to D^{(*)} \bar{D}^{(*)}$ via $D_1\bar{D}$
loop with an effective Lagrangian approach (ELA) under the $D_1 \bar
D$ molecular assumption. The paper is organized as follows. In
Sec.~\ref{sec:formula}, we will introduce the ELA briefly and give
some relevant formulae. In Sec.~\ref{sec:results}, the numerical
results are presented. The summary will be given in
Sec.~\ref{sec:summary}.

\section{The Model}
\label{sec:formula}

%%%%%%%%%%%%%%%%%%%%%%%%%%%%%%%%%%%%%%%%%%%%%%%%%%%%%%%%%%%%%%%%%%%%
\begin{figure}[ht]
\centering
\includegraphics[scale=1.0]{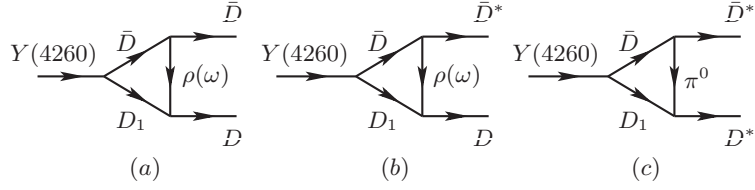}
\caption{The hadron-level diagrams for  $Y(4260) \to
D^{(*)}\bar{D}^{(*)}$ with $D_1{\bar D}$ as the intermediate
states.}\label{fig:feyn-fig1}
\end{figure}

%%%%%%%%%%%%%%%%%%%%%%%%%%%%%%%%%%%%%%%%%%%%%%%%%%%%%%%%%%%%%%%%%%%%
\begin{figure}[ht]
\centering
\includegraphics[scale=1.0]{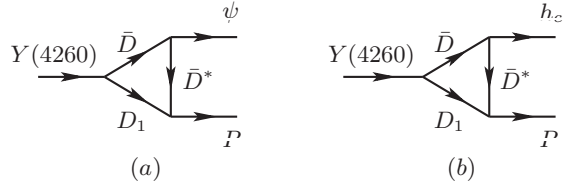}
\caption{The hadron-level diagrams for  hidden-charm decays of
$Y(4260)$ with $D_1{\bar D}$ as intermediate states. $P$
denotes the pseudoscalar meson $\pi^0$ or $\eta$.}\label{fig:feyn-fig2}
\end{figure}
%%%%%%%%%%%%%%%%%%%%%%%%%%%%%%%%%%%%%%%%%%%%%%%%%%%%%%%%%%%%%%%%%%%%

%%%%%%%%%%%%%%%%%%%%%%%%%%%%%%%%%%%%%%%%%%%%%%%%%%%%%%%%%%%%%%%%%%%%
In order to calculate the leading contributions from the charmed
meson loops, we need the leading order effective Lagrangian for the
couplings. Based on the heavy quark symmetry and chiral
symmetry~\cite{Colangelo:2003sa,Casalbuoni:1996pg}, the relevant
effective Lagrangian used in this work read
\begin{eqnarray}
%%psi D(*)D(*)
\mathcal{L}_{\psi D^{(*)} D^{(*)}} &=&
ig_{\psi DD} \psi_{\mu} (\partial^\mu D \bar{D}- D \partial^\mu \bar{D})
-g_{\psi D^* D} \varepsilon^{\mu \nu \alpha \beta}
\partial_{\mu} \psi_{\nu} (\partial_{\alpha} D^*_{\beta} \bar{D} + D \partial_{\alpha}
\bar{D}^*_{\beta})\nonumber\\
&&-ig_{\psi D^* D^*} \big\{
\psi^\mu (\partial_{\mu} D^{* \nu} \bar{D}^*_{\nu}
-D^{* \nu} \partial_{\mu}
\bar{D}^*_{\nu})+ (\partial_{\mu} \psi_{\nu} D^{* \nu} -\psi_{\nu}
\partial_{\mu} D^{* \nu}) \bar{D}^{* \mu} +
D^{* \mu}(\psi^\nu \partial_{\mu} \bar{D}^*_{\nu} -
\partial_{\mu} \psi^\nu \bar{D}^*_{\nu})\big\}, \label{eq:h1}\\
%%
%%hc DD
\mathcal{L}_{h_c D^{(*)} D^{(*)}}&=& g_{h_c D^*
D} h_c^\mu ( D \bar{D}^*_{\mu}+ D^*_\mu \bar{D})+ ig_{h_c
D^* D^*} \varepsilon^{\mu \nu \alpha \beta}
\partial_{\mu} h_{c \nu} D^*_{\alpha} \bar{D}^*_{\beta} \ ,\label{eq:h2}
\end{eqnarray}
where ${{D}^{(*)}}=\left(D^{(*)+},D^{(*)0}, D^{(*)+}_s\right)$ and
${\bar D^{(*)T}}=\left(D^{(*)-},\bar{D}^{(*)0},D_s^{(*)-}\right)$
correspond to the charmed meson isodoublets. The following couplings
are adopted in the numerical calculations,
\begin{eqnarray}
g_{\psi DD} = 2g_2 \sqrt{m_\psi} m_D \ ,
\quad g_{\psi D^* D} = \frac {g_{\psi DD}} {\sqrt{m_D m_{D^*}}} \ ,
\quad g_{\psi D^* D^*} = g_{\psi D^* D}  \sqrt{\frac {m_{D^*}} {m_D}} m_{D^*} \ .
\end{eqnarray}
In principle, the parameter $g_2$ should be computed with
nonperturbative methods. It shows that vector meson dominance (VMD)
would provide an estimate of these
quantities~\cite{Colangelo:2003sa}. The coupling $g_2$ can be
related to the $J/\psi$ leptonic constant $f_\psi$ which is defined
by the matrix element $\langle 0| \bar c\gamma_\mu c|J/\psi (p,
\epsilon)\rangle = f_\psi m_\psi \epsilon^\mu$, and
$g_2={\sqrt{m_\psi}}/2m_Df_\psi,$ where $f_\psi = 405\pm 14$ MeV,
and we have applied the relation $g_{\psi DD} = {m_\psi} /{f_\psi}$.
The ratio of the coupling constants $g_{\psi' DD}$ to $g_{\psi DD}$
is fixed as that in Ref.~\cite{Wang:2012wj}, i.e.,
\begin{eqnarray}
\frac {g_{\psi'DD}} {g_{\psi DD}} = 0.9.
\end{eqnarray}
In addition, the coupling constants in Eq.~(\ref{eq:h2}) are
determined as
\begin{eqnarray}
g_{h_c DD^*} &=& -2g_1 \sqrt{m_{h_c} m_D m_{D^*}} , \ \ g_{h_c
D^* D^*} =2 g_1 \frac{m_{D^*}}{\sqrt{m_{h_c}}},
\end{eqnarray}
with $g_1=-\sqrt{{m_{\chi_{c0}}}/{3}}/{f_{\chi_{c0}}}$, where
$m_{\chi_{c0}}$ and $f_{\chi_{c0}}$ are the mass and decay constant
of $\chi_{c0}(1P)$, respectively~\cite{Colangelo:2002mj}. We take
$f_{\chi_{c0}} = 510 \pm 40$ MeV~\cite{Veliev:2010gb}.

The light vector mesons nonet can be introduced by using the hidden
gauge symmetry approach, and the effective Lagrangian containing
these particles are as
follows~\cite{Casalbuoni:1992gi,Casalbuoni:1992dx},

\begin{eqnarray}
{\cal L}_{{\rm D^*D}V} &=& ig_{{\rm D^*D}V}\epsilon_{\alpha\beta\mu\nu}
({\rm D}_b \stackrel{\leftrightarrow}{\partial_{\alpha}}{\rm D}^{*\beta\dagger}_{a}
-{\rm D}^{*\beta\dagger}_b \stackrel{\leftrightarrow}{\partial_{\alpha}}{\rm D}^{j}_{a})(\partial^\mu V^\nu)_{ba} \nonumber \\
&& +ig_{\overline{\rm D}^* {\overline
D}V}\epsilon_{\alpha\beta\mu\nu} (\overline{\rm D}_b
\stackrel{\leftrightarrow}{\partial_{\alpha}}\overline{\rm
D}^{*\beta\dagger}_{a} -\overline{\rm D}^{*\beta\dagger}_b
\stackrel{\leftrightarrow}{\partial_{\alpha}}\overline{\rm
D}^{j}_{a})(\partial^\mu V^\nu)_{ab}+h.c \ ,
\nonumber \\
{\cal L}_{{\rm DD}V} &=& ig_{{\rm DD}V}({\rm D}_b\stackrel{\leftrightarrow}{\partial_{\mu}}{\rm
D}^{\dagger}_a)V^{\mu}_{ba}+ig_{\overline{\rm D}\,\overline{\rm D}V}(\overline{\rm
D}_b\stackrel{\leftrightarrow}{\partial_{\mu}}\overline{\rm D}^{\dagger}_a)V^{\mu}_{ab} \ , \nonumber \\
{\cal L}_{{\rm DD_1}V}&=&g_{{\rm DD_1}V}{\rm D}^{\mu}_{1b}V_{\mu ba}{\rm D}^{\dagger}_a+g'_{{\rm DD_1}V}({\rm
D}^{\mu}_{1b}\stackrel{\leftrightarrow}{\partial^{\nu}}{\rm D}^{\dagger}_a)(\partial_{\mu}V_{\nu}-\partial_{\nu}V_{\mu})_{ba} \nonumber\\
&&+g_{\overline{\rm D}\,\overline{\rm D}_1 V}\overline{\rm D}^{\dagger}_{a}V_{\mu ab}\overline{\rm
D}^{\mu}_{1b}+g'_{\overline{\rm D}\,\overline{\rm D}_1V}(\overline{\rm
D}^{\mu}_{1b}\stackrel{\leftrightarrow}{\partial^{\nu}}\overline{\rm
D}^{\dagger}_a)(\partial_{\mu}V_{\nu}-\partial_{\nu}V_{\mu})_{ab}+h.c. \ , \nonumber\\
{\cal L}_{{\rm D^{*}D^{*}}V}&=& ig_{{\rm D^{*}D^{*}}V}({\rm
D}^{*}_{b\nu}\stackrel{\leftrightarrow}{\partial_{\mu}}{\rm
D}^{*\nu\dagger}_a)V^{\mu}_{ba}+ig'_{{\rm D^{*}D^{*}}V}({\rm
D}^{*\mu}_{b}{\rm D}^{*\nu\dagger}_{a}-{\rm D}^{*\mu\dagger}_{a}{\rm
D}^{*\nu}_{b})(\partial_{\mu}V_{\nu}-\partial_{\nu}V_{\mu})_{ba} \nonumber\\
&&+ ig_{\overline{\rm D}^{*}\overline{\rm D}^{*}V}(\overline{\rm
D}^{*}_{b\nu}\stackrel{\leftrightarrow}{\partial_{\mu}}\overline{\rm
D}^{*\nu\dagger}_{a})V^{\mu}_{ab}+ig'_{\overline{\rm
D}^{*}\overline{\rm D}^{*}V}(\overline{\rm D}^{*\mu}_b\overline{\rm
D}^{*\nu\dagger}_a-\overline{\rm D}^{*\mu\dagger}_a\overline{\rm
D}^{*\nu}_b)(\partial_{\mu}V_{\nu}-\partial_{\nu}V_{\mu})_{ab} \ .
\end{eqnarray}
And the coupling constants read
\begin{eqnarray}
\nonumber&&g_{{\rm DD}V}=-g_{\overline{\rm D}\,\overline{\rm
D}V}=\frac{1}{\sqrt{2}}\beta g_{V} \ ,\\
\nonumber&&g_{{\rm DD_1}V}=-g_{{\overline{\rm D}\,\overline{\rm
D}_1}V}=-\frac{2}{\sqrt{3}}\,\zeta_1 g_V\sqrt{\rm m_{D}m_{D_1}} \ ,\\
\nonumber&&g'_{{\rm DD_1}V}=-g'_{{\overline{\rm D}\,\overline{\rm
D}_1}V}=\frac{1}{\sqrt{3}}\,\mu_1 g_V  \ ,\\
\nonumber&&g_{{\rm D^{*}D^{*}}V}=-g_{{\overline{\rm
D}^{*}\overline{\rm D}^{*}}V}=-\frac{1}{\sqrt{2}}\,\beta g_{V}  \ ,\\
&&g'_{{\rm D^{*}D^{*}}V}=-g'_{{\overline{\rm D}^{*}\overline{\rm
D}^{*}}V}=-\sqrt{2}\;\lambda g_{V}{\rm m_{D^{*}}} \ ,
\end{eqnarray}
where $f_\pi$ = 132 MeV is the pion decay constant, and the
parameter $g_V$ is given by $g_V = {m_\rho /
f_\pi}$~\cite{Casalbuoni:1996pg}. We take $\lambda = 0.56\,
\text{GeV}^{-1} $, $g = 0.59$ and $\beta=0.9$ in our
calculation~\cite{Isola:2003fh}.

The effective Lagrangian for the light pseudoscalar mesons are
constructed by imposing invariance under both heavy quark
spin-flavor transformation and chiral
transformation~\cite{Burdman:1992gh,Yan:1992gz,Casalbuoni:1996pg,Falk:1992cx}.
The pertinent interaction terms for this work read
\begin{eqnarray}
\mathcal{L}_{D_1D^*P} &=&
g_{D_1D^*P}[3D_{1a}^\mu(\partial_\mu\partial_\nu \phi)_{ab}D^{*\dag
\nu}_{b}
-D_{1a}^\mu(\partial^{\nu}\partial_\nu\phi)_{ab}D_{b\mu}^{*\dag}] \nonumber\\
&+& g_{{\bar D}_1{\bar D}^*P}[3\bar{D}_a^{*\dag
\mu}(\partial_\mu\partial_\nu \phi)_{ab}
\bar{D}_{1b}^\nu-\bar{D}_a^{*\dag \mu}(\partial^\nu \partial_\nu
\phi)_{ab} \bar{D}_{1b\nu}]+H.c. \ , \\
{\cal L}_{\rm DD^{*}P} &=& g_{\rm DD^{*}P}{\rm
D}_b(\partial_{\mu}{\cal \phi})_{ba}{\rm D}^{*\mu\dagger}_a +g_{\rm
DD^{*}P}{\rm D}^{*\mu}_{b}(\partial_{\mu}{\cal \phi})_{ba}{\rm
D}^{\dagger}_a \nonumber\\
 &+&  g_{\rm\overline{D}\,\overline{D}^{*}P}\,\overline{\rm
D}^{*\mu\dagger}_a(\partial_{\mu}{\cal \phi})_{ab}\overline{\rm
D}_b+g_{\rm\overline{D}\,\overline{D}^{*}P}\,\overline{\rm
D}^{\dagger}_a(\partial_{\mu}{\cal \phi})_{ab}\overline{\rm D}^{*\mu}_b
\ ,
\end{eqnarray}
with ${{D}^{(*)}}=\left(D^{(*)+},D^{(*)0}, D^{(*)+}_s\right)$ and
${\bar D^{(*)}}=\left(D^{(*)-},\bar{D}^{(*)0},D_s^{(*)-}\right)$.
$\phi$ is the $3\times 3$ Hermitian matrix for the octet of
Goldstone bosons. In the chiral and heavy quark limit, the above
coupling constants are
\begin{eqnarray}
g_{\rm
DD^{*}P}&=&-g_{\rm\overline{D}\,\overline{D}^{*}P}=-\frac{2g}{f_{\pi}}\sqrt{\rm
m_{\rm D}m_{\rm D^{*}}} \ ,\\
g_{\rm
D^{*}D_1P}&=&g_{\rm\overline{D}^{*}\overline{D}_1P}=-\frac{\sqrt{6}}{3}\,\frac{h^\prime}{\Lambda_{\chi}f_{\pi}}\sqrt{\rm
m_{D^{*}}m_{D_1}} \ ,
\end{eqnarray}
with the chiral symmetry breaking scale $\Lambda_\chi \simeq 1$ GeV
and the coupling $h^\prime = 0.65$~\cite{Deandrea:1999pa}.

By assuming $Y(4260)$ is a $D_1\bar{D}$ molecular state, the
effective Lagrangian is constructed as
\begin{eqnarray}
\mathcal{L}_{Y(4260) D_1D}&=&i\frac {y} {{\sqrt 2}}(\bar{D}_a^\dag Y^\mu
D_{1a}^{\mu\dag}-\bar{D}_{1a}^{\mu\dag} Y^\mu D_a^\dag)+H.c., \label{eq:L-Y}
\end{eqnarray}
which is an $S$-wave coupling. Since the mass $Y(4260)$ is slightly
below an S-wave $D_1\bar D$ threshold, the effective coupling
$g_{Y(4260)D_1 D}$ is related to the probability of finding $D_1D$
component in the physical wave function of the bound state, $c^2$,
and the binding energy, $\delta
E=m_D+m_{D_1}-m_Y$~\cite{Weinberg:1965zz, Baru:2003qq,Guo:2013zbw},
\begin{eqnarray}\label{eq:coupling-Y}
g_{{\rm NR}}^2\equiv 16\pi (m_D+ m_{D_1})^2  c^2\sqrt{\frac {2\delta
E}{\mu}} [1+ {\cal O}(\sqrt{2\mu\epsilon r})]\ ,
\end{eqnarray}
where $\mu=m_Dm_{D_1}/(m_D+m_{D_1})$ and $r$ is the reduced mass and
the range of the forces. The coupling constants in
Eq.~(\ref{eq:L-Y}) is given by the first term in the above equation.
The coupling constant gets maximized for a pure bound state, which
corresponds to $c^2=1$ by definition.  In the following, we present
the numerical results with $c^2=1$.

With the mass $m_Y=4263^{+8}_{-9}$ MeV, and the averaged masses of
the $D$ and $D_1$ mesons~\cite{Beringer:1900zz}, we obtain the mass
differences between the $Y(4260)$ and their corresponding
thresholds,
\begin{eqnarray}
m_D+m_{D_1}-m_Y=27_{-8}^{+9} \ {\rm MeV},
\end{eqnarray}
and with $c^2=1$, we obtain
\begin{eqnarray} \label{eq:coupling-Y1}
|y|=14.62^{+1.11}_{-1.25}\pm 6.20 \ {\rm GeV}
\end{eqnarray}
where the first errors are from the uncertainties of the binding
energies, and the second ones are due the the approximate nature of
the approximate nature of Eq.~(\ref{eq:coupling-Y}).

 The loop transition amplitudes for the transitions in
Figs.~\ref{fig:feyn-fig1} and \ref{fig:feyn-fig2} can be expressed
in a general form in the effective Lagrangian approach as follows,
\begin{eqnarray}
M_{fi}=\int \frac {d^4 q_2} {(2\pi)^4} \sum_{D^* \ \mbox{pol.}}
\frac {T_1T_2T_3} {a_1 a_2 a_3}{\cal F}(m_2,q_2^2)
\end{eqnarray}
where $T_i$ and $a_i = q_i^2-m_i^2 \ (i=1,2,3)$ are the vertex
functions and the denominators of the intermediate meson
propagators. For example, in Fig.~\ref{fig:feyn-fig2} (a), $T_i \
(i=1,2,3)$ are the vertex functions for the initial $Y(4260)$, final
charmonium and final light pseudoscalar mesons, respectively. $a_i \
(i=1,2,3)$  are the denominators for the intermediate $\bar D$,
$D^*$ and $D_1$ mesons, respectively. We introduce a dipole form
factor,
\begin{eqnarray}\label{ELA-form-factor}
{\cal F}(m_{2}, q_2^2) \equiv \left(\frac
{\Lambda^2-m_{2}^2} {\Lambda^2-q_2^2}\right)^2,
\end{eqnarray}
where $\Lambda\equiv m_2+\alpha\Lambda_{\rm QCD}$ and the QCD energy
scale $\Lambda_{\rm QCD} = 220$ MeV. This form factor is supposed to
kill the divergence, compensate the off-shell effects arising from
the intermediate exchanged particle and the non-local effects of the
vertex functions~\cite{Li:1996yn,Locher:1993cc,Li:1996cj}.
\section{Numerical Results}
\label{sec:results}
%%%%%%%%%%%%%%%%%%%%%%%%%%%%%%%%%%%%%%%%%%%%%%%%%%%%%%%%%%%%%%%%%%%%
\begin{figure}[ht]
\centering
\includegraphics[width=0.49\textwidth]{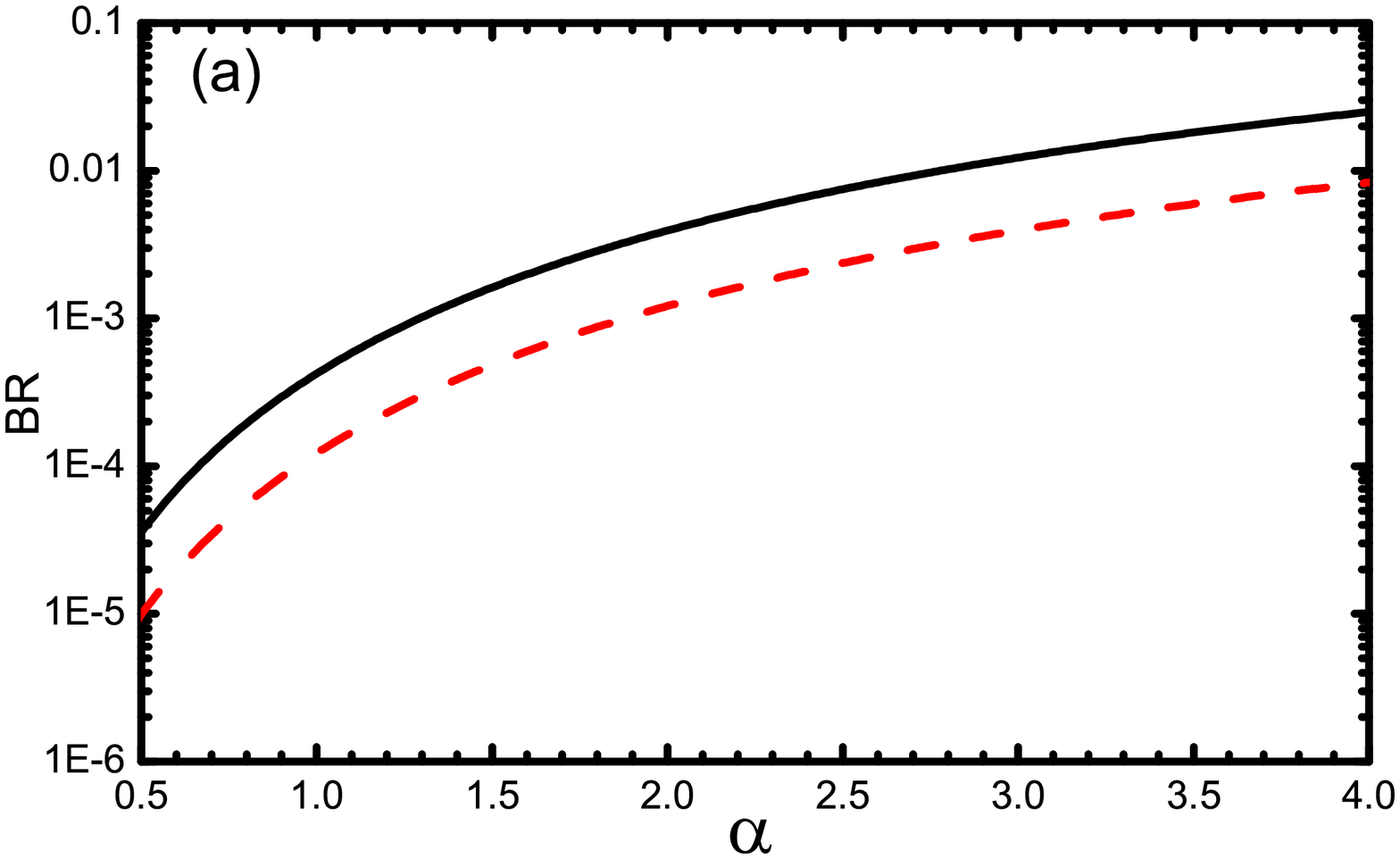}
\includegraphics[width=0.49\textwidth]{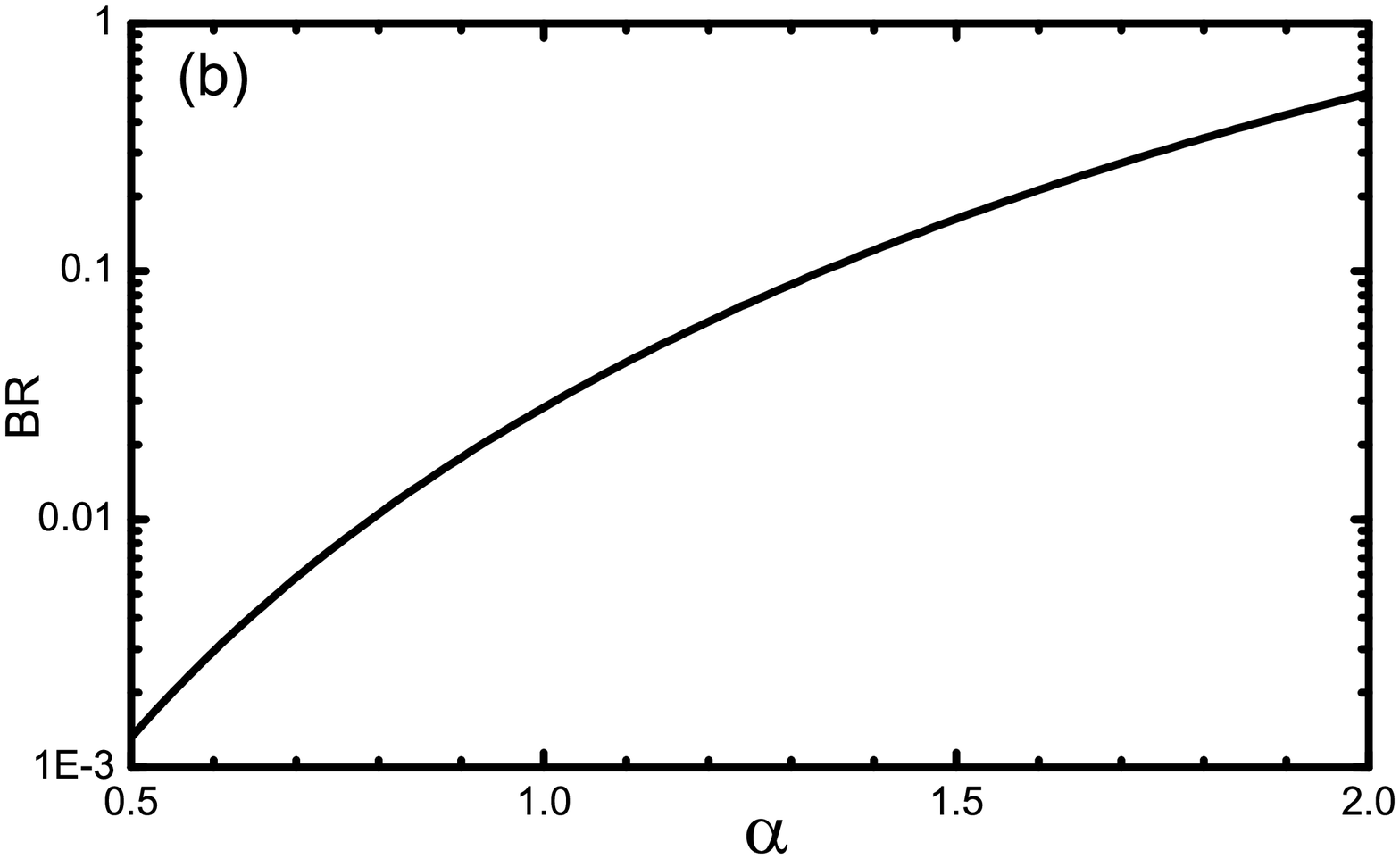}
\caption{(a). The $\alpha$-dependence of the branching ratios of
$Y(4260) \to D {\bar D}$ (solid line) and
$D^{*} {\bar D} + c.c.$ (dashed line). (b).  The $\alpha$-dependence of the branching ratios of
$Y(4260) \to D^{*} {\bar D}^{*}$.}\label{fig:dd-dsd-dsds}
\end{figure}
%%%%%%%%%%%%%%%%%%%%%%%%%%%%%%%%%%%%%%%%%%%%%%%%%%%%%%%%%%%%%%%%%%%%

%%%%%%%%%%%%%%%%%%%%%%%%%%%%%%%%%%%%%%%%%%%%%%%%%%%%%%%%%%%%%%%%%%%%
\begin{figure}[ht]
\centering
\includegraphics[width=0.49\textwidth]{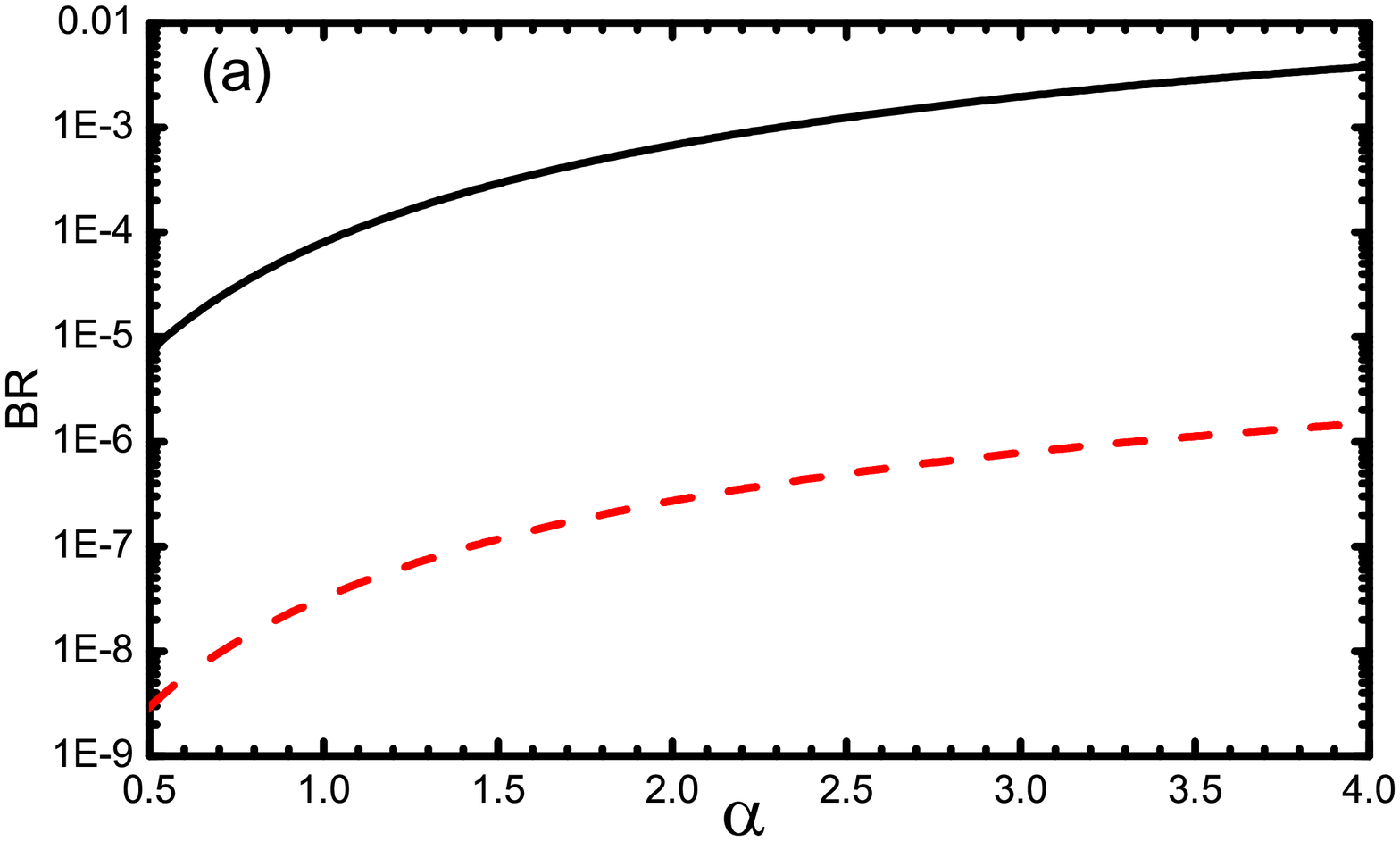}
\includegraphics[width=0.49\textwidth]{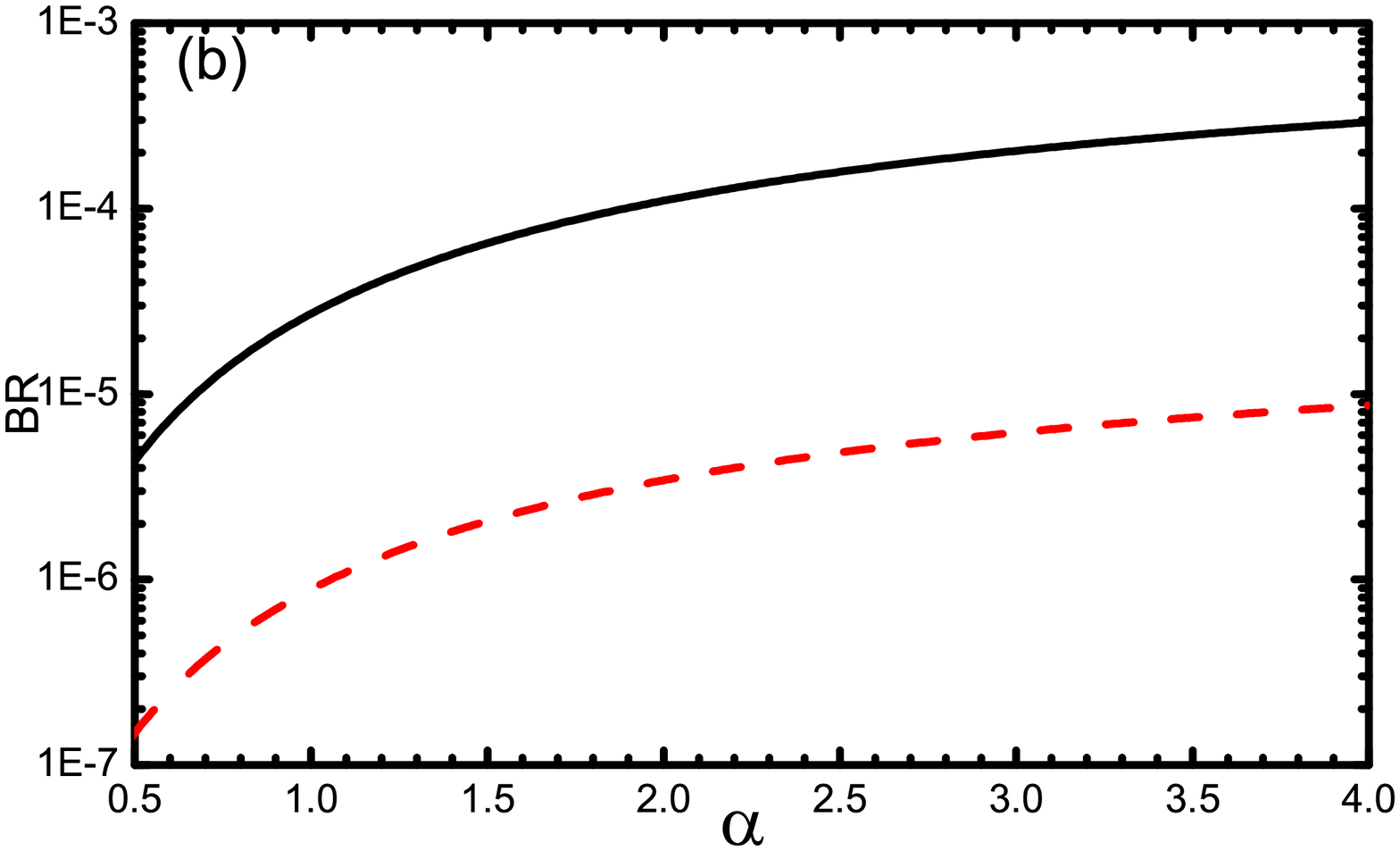}
\caption{(a). The $\alpha$-dependence of the branching ratios of
$Y(4260) \to J/\psi\eta$ (solid line) and
$J/\psi \pi^0$ (dashed line). (b). The $\alpha$-dependence of the branching ratios of
$Y(4260) \to \psi^\prime \eta$ (solid line) and $\psi^\prime \pi^0$ (dashed line).}\label{fig:psi-eta-pi}
\end{figure}
%%%%%%%%%%%%%%%%%%%%%%%%%%%%%%%%%%%%%%%%%%%%%%%%%%%%%%%%%%%%%%%%%%%%

%%%%%%%%%%%%%%%%%%%%%%%%%%%%%%%%%%%%%%%%%%%%%%%%%%%%%%%%%%%%%%%%%%%%
\begin{figure}[ht]
\centering
\includegraphics[width=0.49\textwidth]{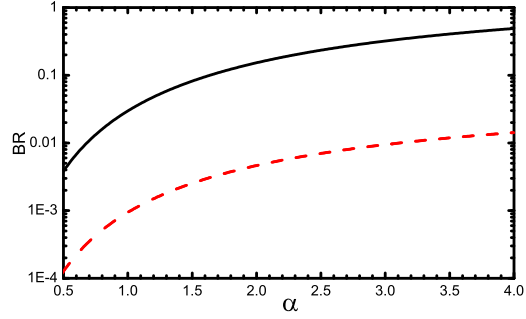}
\caption{The $\alpha$-dependence of the branching ratios of
$Y(4260) \to h_c \eta$ (solid line) and $h_c \pi^0$ (dashed line). }\label{fig:hc-eta-pi}
\end{figure}
%%%%%%%%%%%%%%%%%%%%%%%%%%%%%%%%%%%%%%%%%%%%%%%%%%%%%%%%%%%%%%%%%%%%

%%%%%%%%%%%%%%%%%%%%%%%%%%%%%%%%%%%%%%%%%%%%%%%%%%%%%%%%%%%%%%%%%%%%
\begin{figure}[ht]
\centering
\includegraphics[width=0.49\textwidth]{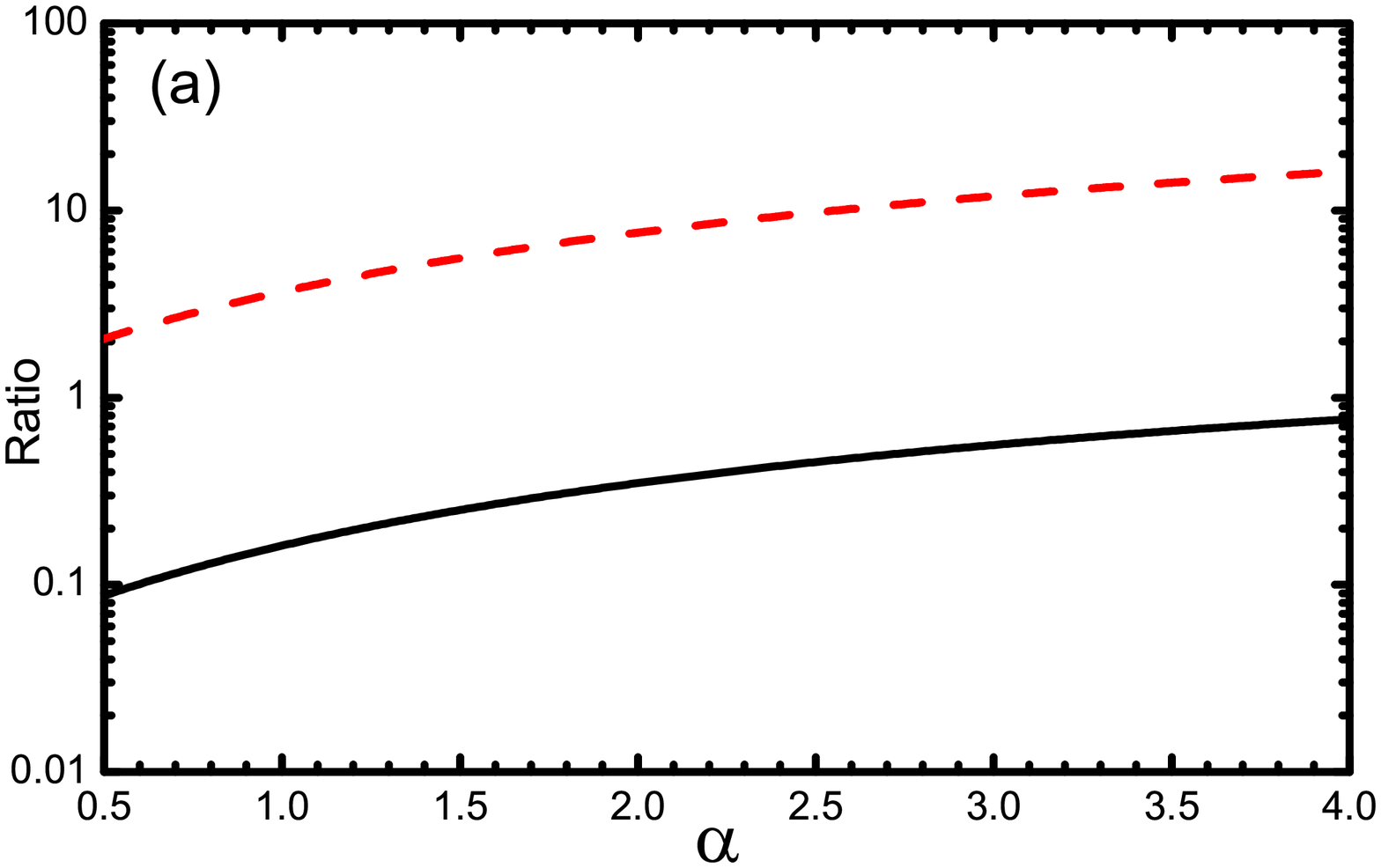}
\includegraphics[width=0.49\textwidth]{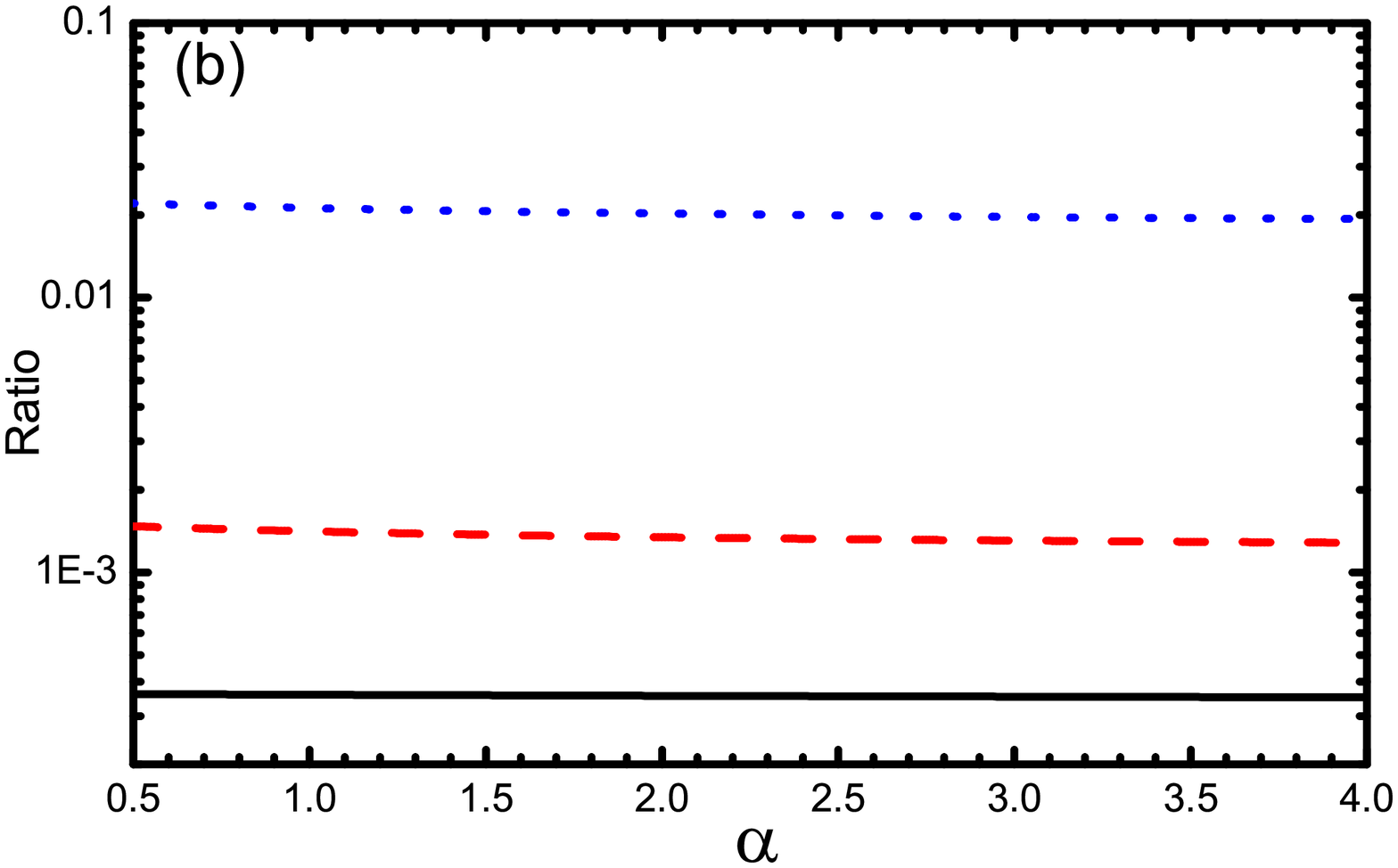}
\caption{(a). The $\alpha$-dependence of the ratios of
$R_1$ (solid line) and $R_2$ (dashed line) defined in Eq.~(\ref{eq:ratio-1}). (b). The $\alpha$-dependence of the ratios of
$r_1$ (solid line), $r_2$ (dashed line) and $r_3$ (dotted line) defined in Eq.~(\ref{eq:ratio-2}). }\label{fig:ratio-psi}
\end{figure}
%%%%%%%%%%%%%%%%%%%%%%%%%%%%%%%%%%%%%%%%%%%%%%%%%%%%%%%%%%%%%%%%%%%%

\begin{table}
\begin{center}
\caption{The predicted branching ratios of Y(4260) decays with
different $\alpha$ values. The uncertainties is dominated by the use
of Eq.~(\ref{eq:coupling-Y}).}\label{tab:br}
\begin{tabular}{ccccc}
\hline
Final states       & $\alpha =0.5$ & $\alpha=1.0 $& $\alpha=1.5 $   & $\alpha=2.0 $ \\
\hline
$D {\bar D}$  & $(3.54^{+3.71}_{-2.34})\times 10^{-5}$& $(4.21^{+4.41}_{-2.78})\times 10^{-4}$ & $(1.62^{+1.70}_{-1.07})\times 10^{-3}$ & $(3.94^{+4.13}_{-2.60})\times 10^{-3}$ \\
$D^{*}{\bar D} + c.c.$ & $(9.86^{+10.33}_{-6.51})\times 10^{-6}$& $(1.22^{+1.28}_{-0.80})\times 10^{-4}$ & $(4.82^{+5.05}_{-3.18})\times 10^{-4}$ & $(1.20^{+1.28}_{-0.79})\times 10^{-3}$ \\
$D^{*}{\bar D}^{*}$ & $(1.41^{+1.48}_{-0.93})\times 10^{-3}$& $(2.78^{+2.91}_{-1.83})\times 10^{-2}$ & $(16.24^{+17.01}_{-10.72})\%$ & $(52.21^{+54.69}_{-34.48})\%$ \\
$J/\psi \eta$ & $(7.43^{+7.78}_{-4.91})\times 10^{-6}$& $(8.19^{+8.58}_{-5.41})\times 10^{-5}$ & $(2.95^{+3.09}_{-1.95})\times 10^{-4}$ & $(6.80^{+7.12}_{-4.49})\times 10^{-4}$ \\
$J/\psi \pi^0$ & $(3.04^{+3.18}_{-2.01})\times 10^{-9}$& $(3.32^{+3.48}_{-2.19})\times 10^{-8}$ & $(1.19^{+1.24}_{-0.78})\times 10^{-7}$ & $(2.72^{+2.85}_{-1.79})\times 10^{-7}$ \\
$\psi^\prime \eta$ & $(4.34^{+4.54}_{-2.84})\times 10^{-6}$& $(2.71^{+2.84}_{-1.79})\times 10^{-5}$ & $(6.50^{+6.81}_{-4.29})\times 10^{-5}$ & $(1.10^{+1.15}_{-0.73})\times 10^{-4}$ \\
$\psi^\prime \pi^0$ & $(1.76^{+1.84}_{-1.16})\times 10^{-7}$& $(9.71^{+10.17}_{-6.41})\times 10^{-7}$ & $(2.14^{+2.24}_{-1.41})\times 10^{-6}$ & $(3.43^{+3.59}_{-2.26})\times 10^{-6}$ \\
$h_c \eta$ & $(3.87^{+4.05}_{-2.55})\times 10^{-3}$& $(2.99^{+3.13}_{-1.97})\times 10^{-2}$ & $(8.20^{+8.59}_{-5.41})\times 10^{-2}$ & $(15.26^{+15.98}_{-10.08})\%$ \\
$h_c \pi^0$ & $(1.27^{+1.33}_{-0.84})\times 10^{-4}$& $(9.50^{+9.95}_{-6.27})\times 10^{-4}$ & $(2.54^{+2.66}_{-1.67})\times 10^{-3}$ & $(4.62^{+4.83}_{-3.05})\times 10^{-3}$ \\
\hline
\end{tabular}
\end{center}
\end{table}
Since $Y(4260)$ has a large width $95\pm 14$ MeV, so one has to take
into account the mass distribution of the $Y(4260)$ when calculating
its decay widths. Its two-body decay width can then be calculated as
follow~\cite{Cleven:2011gp},
\begin{eqnarray}
\Gamma (Y(4260))_{\rm 2-body}=\frac {1} {W} \int_{(m_Y-2\Gamma_Y)^2}^{(m_Y+2\Gamma_Y)^2}ds
\frac {(2\pi)^4}{2{\sqrt s}} \int d\Phi_2 |{\cal M}|^2 \frac {1}{\pi} {\rm Im} (\frac {-1} {s-m_Y^2 +im_Y \Gamma_Y})
\end{eqnarray}
$\int d\Phi_2$ is the two-body phase space~\cite{Beringer:1900zz}.
$\cal M$ are the loop transition amplitudes for the processes in
Figs.~\ref{fig:feyn-fig1} and \ref{fig:feyn-fig2}. The factor $1/W$
with
\begin{eqnarray}
W= \frac {1}{\pi} \int_{(m_Y-2\Gamma_Y)^2}^{(m_Y+2\Gamma_Y)^2}
{\rm Im} (\frac {-1} {s-m_Y^2 +im_Y \Gamma_Y})ds
\end{eqnarray}
is considered in order to normalize the spectral function of the
$Y(4260)$ state.

The numerical results are presented in
Figs.~\ref{fig:dd-dsd-dsds}-\ref{fig:hc-eta-pi}. In
Table.~\ref{tab:br}, we list the predicted branching ratios of
$Y(4260)$ at different $\alpha$ values and the errors are from the
uncertainties of the the coupling constants in
Eq.~(\ref{eq:coupling-Y1}). We have checked that including the width
for the $D_1$ only causes a minor change of about 1\%-3\%.

In Fig.~\ref{fig:dd-dsd-dsds}(a), we plot the $\alpha$-dependence of
the branching ratios of $Y(4260) \to D {\bar D}$ (solid line) and
$Y(4260) \to D^{*} {\bar D} + c.c.$ (dashed line), respectively. The
branching ratios of $Y(4260) \to D^{*} {\bar D}^{*}$ in terms of
$\alpha$ are shown in Fig.~\ref{fig:dd-dsd-dsds}(b). In this figure,
no cusp structure appear. This is because that the mass of $Y(4260)$
lies below the intermediate $D_1\bar D$ threshold. The $\alpha$
dependence of the branching ratios are not drastically sensitive to
some extent, which indicates a reasonable cutoff of the ultraviolet
contributions by the empirical form factors. As shown in this
figure, at the same $\alpha$, the intermediate $D_1 {\bar D}$ meson
loops turns out to be more important in $Y(4260) \to D^{*} {\bar
D}^{*}$ than that in $Y(4260) \to D{\bar D}$ and $D^{*} {\bar D}
+c.c.$. This behavior can also be seen from Table.~\ref{tab:br}. As
a result, a smaller value of $\alpha$ is favored in $Y(4260) \to
D^{*}{\bar D}^{*}$. This phenomenon can be easily explained from
Fig.~\ref{fig:feyn-fig1}. For the decay $Y(4260) \to D^{*} {\bar
D}^{*}$, the off-shell effects of intermediate mesons $D_1D(\pi)$
are not significant, which makes this decay favor a relatively
smaller $\alpha$ value. For the decay $Y(4260) \to D{\bar D}$ and
$D^{*} {\bar D} +c.c.$, since the exchanged mesons of the
intermediate meson loops are $\rho$ and $\omega$, which makes their
off-effects are relatively significant, which makes this decay favor
a relatively larger $\alpha$ value.

In a fit to the total hadronic cross sections measured by
BES~\cite{Bai:2001ct}, authors set an upper limit on
$\Gamma_{e^+e^-}$ for $Y(4260)$ to be less than $580 \, {\rm eV}$ at
90\% confidence level (C.L.)~\cite{Mo:2006ss}. This implies that its
branching fraction to $J/\psi \pi^+\pi^-$ is greater than 0.6\% at
90\% C.L.~\cite{Mo:2006ss}. Recently, BESIII has reported a study of
$e^+e^- \to h_c\pi^+\pi^-$, and observes a state with a mass of
$4021.8\pm 1.0\pm 2.5 \,, {\rm MeV}$ and a width of $5.7\pm 3.4\pm
1.1 {\rm MeV}$ in the $h_c\pi^{\pm}$ mass distribution, called the
$Z_c(4020)$. The Belle collaboration did a comprehensive search for
$Y(4260)$ decays to all possible final states containing open
charmed mesons pairs and found no sign of a $Y(4260)$ signal in any
of
them~\cite{Pakhlova:2010ek,Pakhlova:2008vn,Pakhlova:2009jv,Pakhlova:2007fq,Abe:2006fj,Pakhlova:2008zza}.
The BaBar Collaboration measured some upper limits of the ratios
${{\cal B} (Y(4260) \to D\bar D)}/{{\cal B} (Y(4260)\to J/\psi
\pi^+\pi^-)}<7.6$ at $95\%$ C.L.~\cite{Aubert:2006mi}, ${{\cal B}
(Y(4260) \to D^*\bar D)}/{{\cal B} (Y(4260)\to J/\psi \pi^+\pi^-)} <
34$ and ${{\cal B} (Y(4260) \to D^*{\bar D}^*)}/{{\cal B}
(Y(4260)\to J/\psi \pi^+\pi^-)} < 40$ at 90\%
C.L.~\cite{Aubert:2009aq}, respectively. Within the parameter range
considered in this work, the results displayed in Table~\ref{tab:br}
could be compatible with these available experimental limits.
However, since there are still several uncertainties coming from the
undetermined coupling constants, and the cutoff energy dependence of
the amplitude is not quite stable, the numerical results would be
lacking in high accuracy. Especially, since the kinematics,
off-shell effects arising from the exchanged particles and the
divergence of the loops in theses open charmed channels studied here
are different, the cutoff parameter can also be different in
different decay channels. We expect more precise experimental
measurements on these open charmed pairs to test this point in the
near future.

In Ref.~\cite{Guo:2009wr}, a nonrelativistic effective field theory
(NREFT) method was introduced to study the meson loop effects in
$\psi^\prime \to J/\psi \pi^0$ transitions. And a power counting
scheme was proposed to estimate the contribution of the loop
effects, which is helpful to judge how important the coupled-channel
effects are. This power counting scheme was analyzed in detail in
Ref.~\cite{Guo:2010ak}. Recently, the authors study that the S-wave
threshold plays more important role than P-wave, especially for the
S-wave molecule with large coupling to its components, such as
$Y(4260)$ coupling to $D_1 \bar{D}$ in Ref.~\cite{Guo:2013zbw}.
Before giving the explicit numerical results, we will follow the
similar power counting scheme to qualitatively estimate the
contributions of the coupled-channel effects discussed in this work.
Corresponding to the diagrams Fig.~\ref{fig:feyn-fig2}(a) and
Fig.~\ref{fig:feyn-fig2}(b), the amplitudes for $Y(4260)\to
J/\psi\pi^0$ ($J/\psi\eta$, $\psi^\prime \pi^0$, $\psi^\prime \eta$)
and $Y(4260)\to h_c\pi^0$ ($h_c\eta$) scale as
\begin{eqnarray}
\frac{v^5}{(v^2)^3} q^3 \frac{\Delta}{v^2}\sim
\frac{q^3\Delta}{v^3} \ \label{power:jpsipi}
\end{eqnarray}
and
\begin{eqnarray}
\frac{v^5}{(v^2)^3} q^2 \frac{\Delta}{v^2}\sim \frac{q^2\Delta}{v^3}
\ , \label{power:hcpi}
\end{eqnarray}
respectively. There are two scaling parameters $v$ and $q$ appeared
in the above two formulae. As illustrated in Ref.~\cite{Guo:2012tg},
$v$ is understood as the average velocity of the intermediated
charmed meson. $q$ denotes the momentum of the outgoing pseudoscalar
meson. And $\Delta$ denotes the charmed meson mass difference, which
is introduced to account for the isospin or SU(3) symmetry
violation. For the $\pi^0$ and $\eta$ production processes, the
factors $\Delta$ are about $M_{D^+}+M_{D^-}-2M_{D^0}$ and
$M_{D^+}+M_{D^0}-2M_{D_s}$, respectively. According to
Eqs.~(\ref{power:jpsipi}) and (\ref{power:hcpi}), it can be
concluded that the contributions of the coupled channel effects
would be significant here since the amplitudes scale as
$\mathcal{O}(1/v^3)$. And the branching ratio of $Y(4260)\to
h_c\pi^0$ is expected to be larger than that of $Y(4260)\to
J/\psi\pi^0$, because the corresponding amplitudes scale as
$\mathcal{O}(q^2)$ and $\mathcal{O}(q^3)$ respectively. However, the
momentum $q$ in $Y(4260)\to J/\psi\pi^0$ is larger than that in
$Y(4260)\to h_c\pi^0$, which may compensate this discrepancy to some
extent.

For the open charmed decays in Fig.~\ref{fig:feyn-fig1}, the
exchanged intermediate mesons are light vector mesons or light
pseudoscalar mesons which will introduce different scale. Since we
cannot separate different scales, so we just give possible numerical
results in the form factor scheme.

For the hidden-charm transitions $Y(4260) \to J/\psi\eta (\pi^0)$,
we plot the $\alpha$-dependence of the branching ratios of $Y(4260)
\to J/\psi\eta (\pi^0)$ in Fig.~\ref{fig:psi-eta-pi}(a) as shown by
the solid and dashed lines, respectively. The $\pi^0$-$\eta$ mixing
has been taken into account. (Using Dashen's
theorem~\cite{Dashen:1969eg}, one may express the mixing angle in
terms of the masses of the Goldstone bosons at leading order in
chiral perturbation theory and the value is about 0.01). Some points
can be learned from this figure: (1). A predominant feature is that
the branching ratios are not drastically sensitive to the cutoff
parameter, which indicates a reasonable cutoff of the ultraviolet
contributions by the empirical form factors to some extent. (2). The
leading contributions to the $Y(4260) \to J/\psi\pi^0$ are given by
the differences between the neutral and charged charmed meson loops
and also from the $\pi^0$-$\eta$ mixing through the loops
contributing to the eta transition. (3). At the same $\alpha$, the
branching ratios for $Y(4260) \to J/\psi \eta$ transition are 2-3
orders of magnitude larger than that of $Y(4260) \to J/\psi \pi^0$.
It is because that there is no cancelations between the charged and
neutral meson loops.

The branching ratios of $Y(4260) \to \psi^\prime \eta$ (solid line)
and $Y(4260) \to \psi^\prime \pi^0$ (dashed line) in terms of
$\alpha$ are shown in Fig.~\ref{fig:psi-eta-pi}(b). The behavior is
similar to that of Fig.~\ref{fig:psi-eta-pi}(a). Since the mass of
$\psi^\prime$ is closer to the thresholds of $\bar{D}D^*$ than
$J/\psi$, it should give rise to important threshold effects in
$Y(4260) \to \psi^\prime \eta (\pi^0)$ than in $Y(4260) \to J/\psi
\eta (\pi^0)$. At the same $\alpha$ value, the obtained branching
ratios of $Y(4260) \to \psi^\prime \pi^0$ is larger than that of
$Y(4260) \to J/\psi\pi^0$. Since the three-momentum of final $\eta$
is only about $167$ MeV in $Y(4260) \to \psi^\prime \eta$, which
lead to a smaller branching rations in $Y(4260) \to J/\psi\eta$ than
that in $Y(4260) \to J/\psi\eta$ at the same $\alpha$ value.

In Fig.~\ref{fig:hc-eta-pi}, we plot the $\alpha$-dependence of the
branching ratios of $Y(4260) \to h_c \pi^0$ (solid line) and
$Y(4260) \to h_c \eta$ (dashed line), respectively. The branching
ratios for $Y(4260) \to h_c\pi^0(\eta)$ are larger than that of
$Y(4260) \to J/\psi\pi^0(\eta)$ and $\psi^\prime\pi^0(\eta)$, which
is consistent with the power counting analysis in
Eqs.~(\ref{power:jpsipi}) and (\ref{power:hcpi}).

In order to study the exclusive threshold effects via the
intermediate mesons loops, we define the following ratios,

\begin{eqnarray}
R_1&\equiv& \frac {|{\cal M}_{Y(4260) \to \psi^\prime \pi}|^2 } {|{\cal M}_{Y(4260) \to J/\psi \pi}|^2}, \quad
R_2\equiv \frac {|{\cal M}_{Y(4260) \to \psi^\prime \eta}|^2} {|{\cal M}_{Y(4260) \to J/\psi \eta}|^2}  ,\label{eq:ratio-1}
\end{eqnarray}
and
\begin{eqnarray}
r_1&\equiv& \frac {|{\cal M}_{Y(4260) \to J/\psi \pi}|^2} {|{\cal M}_{Y(4260) \to J/\psi \eta}|^2}  , \quad
r_2\equiv \frac {|{\cal M}_{Y(4260) \to \psi^\prime \pi}|^2} {|{\cal M}_{Y(4260) \to \psi^\prime \eta}|^2} , \quad
r_3\equiv \frac {|{\cal M}_{Y(4260) \to h_c \pi}|^2} {|{\cal M}_{Y(4260) \to h_c \eta}|^2} . \label{eq:ratio-2}
\end{eqnarray}
These ratios are plotted in Fig.~\ref{fig:ratio-psi}(a) and (b),
respectively. The stabilities of the ratios in terms of $\alpha$
indicate a reasonably controlled cutoff for each channels by the
form factor. Since the coupling vertices are the same for those
decay channels when taking the ratio, the stability of the ratios
suggests that the transitions of $Y(4260) \to J/\psi\pi^0 (\eta)$
and $\psi^\prime \pi^0(\eta)$ are largely driven by the open
threshold effects via the intermediate $D_1\bar D$ meson loops to
some extent. The future experimental measurements of these decays
can help us investigate this issue deeply.

\section{Summary}
\label{sec:summary}

In this work, we have investigated the hidden-charm decays of
$Y(4260)$ and  the decays $Y(4260)\to D{\bar D}$, $D {\bar D}^*$ and
$D^*{\bar D}^*$ in ELA. In this calculation, $Y(4260)$ is assumed to
be the $D_1 {\bar D}$ molecular state. Our results show that the
$\alpha$ dependence of the branching ratios are not drastically
sensitive, which indicate the dominant mechanism driven by the
intermediate meson loops with a fairly well control of the
ultraviolet contributions.

For the hidden charmonium decays, we also carried out the power
counting analysis and our results for these decays in ELA are
qualitatively consistent with the power counting analysis. For the
open charmed decays $Y(4260) \to D\bar D$, $D {\bar D}^*$ and
$D^*{\bar D}^*$, the exchanged intermediate mesons are light vector
mesons or light pseudoscalar mesons which will introduce different
scale, so we cannot separate different scales and only give possible
numerical results in the form factor scheme. For the decay $Y(4260)
\to D^{*}{\bar D}^{*}$, the exchanged mesons $\pi$ is almost
on-shell, so the coupled channel effects are more important than
other channels studied here. We expect the experiments to search for
the hidden-charm and charmed meson pairs decays of $Y(4260)$, which
will help us investigate the nature and decay mechanisms of
$Y(4260)$ deeply.

\section*{Acknowlegements}

Authors thank Prof. Q. Zhao, Q. Wang and D.-Y. Chen for useful
discussions. This work is supported in part by the National Natural
Science Foundation of China (Grant Nos. 11035006, and 11275113), and
in part by the China Postdoctoral Science Foundation (Grant No.
2013M530461).


\begin{thebibliography}{99}
%\cite{Swanson:2006st}
\bibitem{Swanson:2006st}
  E.~S.~Swanson,
  %``The New heavy mesons: A Status report,''
  Phys.\ Rept.\  {\bf 429}, 243 (2006)  [hep-ph/0601110].
  %%CITATION = HEP-PH/0601110;%%  %416 citations counted in INSPIRE as of 08 Jul 2013

%\cite{Eichten:2007qx}
\bibitem{Eichten:2007qx}
  E.~Eichten, S.~Godfrey, H.~Mahlke and J.~L.~Rosner,
  %``Quarkonia and their transitions,''
  Rev.\ Mod.\ Phys.\  {\bf 80}, 1161 (2008)  [hep-ph/0701208].
  %%CITATION = HEP-PH/0701208;%%  %154 citations counted in INSPIRE as of 08 Jul 2013

%\cite{Voloshin:2007dx}
\bibitem{Voloshin:2007dx}
  M.~B.~Voloshin,
  %``Charmonium,''
  Prog.\ Part.\ Nucl.\ Phys.\  {\bf 61}, 455 (2008)  [arXiv:0711.4556 [hep-ph]].
  %%CITATION = ARXIV:0711.4556;%%  %161 citations counted in INSPIRE as of 08 Jul 2013


%\cite{Godfrey:2008nc}
\bibitem{Godfrey:2008nc}
  S.~Godfrey and S.~L.~Olsen,
  %``The Exotic XYZ Charmonium-like Mesons,''
  Ann.\ Rev.\ Nucl.\ Part.\ Sci.\  {\bf 58}, 51 (2008)  [arXiv:0801.3867 [hep-ph]].
  %%CITATION = ARXIV:0801.3867;%%  %148 citations counted in INSPIRE as of 08 Jul 2013

%\cite{Drenska:2010kg}
\bibitem{Drenska:2010kg}
  N.~Drenska, R.~Faccini, F.~Piccinini, A.~Polosa, F.~Renga and C.~Sabelli,
  %``New Hadronic Spectroscopy,''
  Riv.\ Nuovo Cim.\  {\bf 033}, 633 (2010)  [arXiv:1006.2741 [hep-ph]].
  %%CITATION = ARXIV:1006.2741;%%  %39 citations counted in INSPIRE as of 08 Jul 2013


%\cite{Aubert:2005rm}
\bibitem{Aubert:2005rm}
  B.~Aubert {\it et al.}  [BaBar Collaboration],
  %``Observation of a broad structure in the $\pi^+ \pi^- J/\psi$ mass spectrum around 4.26-GeV/c$^2$,''
  Phys.\ Rev.\ Lett.\  {\bf 95}, 142001 (2005)  [hep-ex/0506081].
  %%CITATION = HEP-EX/0506081;%%  %448 citations counted in INSPIRE as of 23 Apr 2013

%\cite{He:2006kg}
\bibitem{He:2006kg}
  Q.~He {\it et al.}  [CLEO Collaboration],
  %``Confirmation of the Y(4260) resonance production in ISR,''
  Phys.\ Rev.\ D {\bf 74}, 091104 (2006)  [hep-ex/0611021].
  %%CITATION = HEP-EX/0611021;%%  %106 citations counted in INSPIRE as of 23 Apr 2013

%\cite{Yuan:2007sj}
\bibitem{Yuan:2007sj}
  C.~Z.~Yuan {\it et al.}  [Belle Collaboration],
  %``Measurement of e+ e- ---> pi+ pi- J/psi cross-section via initial state radiation at Belle,''
  Phys.\ Rev.\ Lett.\  {\bf 99}, 182004 (2007)  [arXiv:0707.2541 [hep-ex]].
  %%CITATION = ARXIV:0707.2541;%%  %195 citations counted in INSPIRE as of 23 Apr 2013

\bibitem{Beringer:1900zz}
  J.~Beringer {\it et al.}  [Particle Data Group Collaboration],
  %``Review of Particle Physics (RPP),''
  Phys.\ Rev.\ D {\bf 86}, 010001 (2012).
  %%CITATION = PHRVA,D86,010001;%%  %997 citations counted in INSPIRE as of 12 Mar 2013

%\cite{Ablikim:2013mio}
\bibitem{Ablikim:2013mio}
  M.~Ablikim {\it et al.}  [BESIII Collaboration],
  %``Observation of a charged charmoniumlike structure in $e^+e^- \to \pi^+\pi^-J/\psi$ at $\sqrt{s}=4.26$ GeV,''
  Phys.\ Rev.\ Lett.\  {\bf 110}, 252001 (2013)  [arXiv:1303.5949 [hep-ex]].
  %%CITATION = ARXIV:1303.5949;%%  %36 citations counted in INSPIRE as of 07 Jul 2013


%\cite{Brambilla:2010cs}
\bibitem{Brambilla:2010cs}
  N.~Brambilla, S.~Eidelman, B.~K.~Heltsley, R.~Vogt, G.~T.~Bodwin, E.~Eichten, A.~D.~Frawley and A.~B.~Meyer {\it et al.},
  %``Heavy quarkonium: progress, puzzles, and opportunities,''
  Eur.\ Phys.\ J.\ C {\bf 71}, 1534 (2011)  [arXiv:1010.5827 [hep-ph]].
  %%CITATION = ARXIV:1010.5827;%%  %335 citations counted in INSPIRE as of 12 Apr 2013

%\cite{LlanesEstrada:2005hz}
\bibitem{LlanesEstrada:2005hz}
  F.~J.~Llanes-Estrada,
  %``Y(4260) and possible charmonium assignment,''
  Phys.\ Rev.\ D {\bf 72}, 031503 (2005)  [hep-ph/0507035].
  %%CITATION = HEP-PH/0507035;%%  %54 citations counted in INSPIRE as of 30 Apr 2013

%\cite{Maiani:2005pe}
\bibitem{Maiani:2005pe}
  L.~Maiani, V.~Riquer, F.~Piccinini and A.~D.~Polosa,
  %``Four quark interpretation of Y(4260),''
  Phys.\ Rev.\ D {\bf 72}, 031502 (2005)  [hep-ph/0507062].
  %%CITATION = HEP-PH/0507062;%%  %162 citations counted in INSPIRE as of 30 Apr 2013

%\cite{Zhu:2005hp}
\bibitem{Zhu:2005hp}
  S.~-L.~Zhu,
  %``The Possible interpretations of Y(4260),''
  Phys.\ Lett.\ B {\bf 625}, 212 (2005)  [hep-ph/0507025].
  %%CITATION = HEP-PH/0507025;%%  %115 citations counted in INSPIRE as of 30 Apr 2013

%\cite{Kou:2005gt}
\bibitem{Kou:2005gt}
  E.~Kou and O.~Pene,
  %``Suppressed decay into open charm for the Y(4260) being an hybrid,''
  Phys.\ Lett.\ B {\bf 631}, 164 (2005)  [hep-ph/0507119].
  %%CITATION = HEP-PH/0507119;%%  %115 citations counted in INSPIRE as of 30 Apr 2013

%\cite{Close:2005iz}
\bibitem{Close:2005iz}
  F.~E.~Close and P.~R.~Page,
  %``Gluonic charmonium resonances at BaBar and BELLE?,''
  Phys.\ Lett.\ B {\bf 628}, 215 (2005)  [hep-ph/0507199].
  %%CITATION = HEP-PH/0507199;%%  %147 citations counted in INSPIRE as of 30 Apr 2013

%\cite{Ding:2007rg}
\bibitem{Ding:2007rg}
  G.~-J.~Ding, J.~-J.~Zhu and M.~-L.~Yan,
  %``Canonical Charmonium Interpretation for Y(4360) and Y(4660),''
  Phys.\ Rev.\ D {\bf 77}, 014033 (2008)  [arXiv:0708.3712 [hep-ph]].
  %%CITATION = ARXIV:0708.3712;%%  %38 citations counted in INSPIRE as of 30 Apr 2013

%\cite{Ding:2008gr}
\bibitem{Ding:2008gr}
  G.~-J.~Ding,
  %``Are Y(4260) and Z+(2) are D(1) D or D(0) D* Hadronic Molecules?,''
  Phys.\ Rev.\ D {\bf 79}, 014001 (2009)  [arXiv:0809.4818 [hep-ph]].
  %%CITATION = ARXIV:0809.4818;%%  %45 citations counted in INSPIRE as of 30 Apr 2013

%\cite{Wang:2013cya}
\bibitem{Wang:2013cya}
  Q.~Wang, C.~Hanhart and Q.~Zhao,
  %``Decoding the riddle of Y(4260) and $Z_c(3900)$,''
  arXiv:1303.6355 [hep-ph].
  %%CITATION = ARXIV:1303.6355;%%  %11 citations counted in INSPIRE as of 30 Apr 2013

%\cite{Filin:2010se}
\bibitem{Filin:2010se}
  A.~A.~Filin, A.~Romanov, V.~Baru, C.~Hanhart, Y.~.S.~Kalashnikova, A.~E.~Kudryavtsev, U.~-G.~Mei{\ss}ner and A.~V.~Nefediev,
  %``Comment on `Possibility of Deeply Bound Hadronic Molecules from Single Pion Exchange',''
  Phys.\ Rev.\ Lett.\  {\bf 105}, 019101 (2010)  [arXiv:1004.4789 [hep-ph]].
  %%CITATION = ARXIV:1004.4789;%%  %7 citations counted in INSPIRE as of 23 Aug 2013

%\cite{Guo:2011dd}
\bibitem{Guo:2011dd}
  F.~-K.~Guo and U.~-G.~Mei{\ss}ner,
  %``More kaonic bound states and a comprehensive interpretation of the $D_{sJ}$ states,''
  Phys.\ Rev.\ D {\bf 84}, 014013 (2011)  [arXiv:1102.3536 [hep-ph]].
  %%CITATION = ARXIV:1102.3536;%%  %5 citations counted in INSPIRE as of 23 Aug 2013

%\cite{Guo:2013zbw}
\bibitem{Guo:2013zbw}
  F.~-K.~Guo, C.~Hanhart, U.~-G.~Mei{\ss}ner, Q.~Wang and Q.~Zhao,
  %``Production of the X(3872) in charmonia radiative decays,''
  Phys.\  Lett.\  B {\bf 725}, 127 (2013)  [arXiv:1306.3096 [hep-ph]].
  %%CITATION = ARXIV:1306.3096;%%  %2 citations counted in INSPIRE as of 09 Aug 2013

%\cite{Yuan:2005dr}
\bibitem{Yuan:2005dr}
  C.~Z.~Yuan, P.~Wang and X.~H.~Mo,
  %``The Y(4260) as an omega chi(c1) molecular state,''
  Phys.\ Lett.\ B {\bf 634}, 399 (2006)  [hep-ph/0511107].
  %%CITATION = HEP-PH/0511107;%%  %29 citations counted in INSPIRE as of 04 May 2013

%\cite{Liu:2005ay}
\bibitem{Liu:2005ay}
  X.~Liu, X.~-Q.~Zeng and X.~-Q.~Li,
  %``Possible molecular structure of the newly observed Y(4260),''
  Phys.\ Rev.\ D {\bf 72}, 054023 (2005)  [hep-ph/0507177].
  %%CITATION = HEP-PH/0507177;%%  %69 citations counted in INSPIRE as of 04 May 2013

%\cite{MartinezTorres:2009xb}
\bibitem{MartinezTorres:2009xb}
  A.~Martinez Torres, K.~P.~Khemchandani, D.~Gamermann and E.~Oset,
  %``The Y(4260) as a J/psi K anti-K system,''
  Phys.\ Rev.\ D {\bf 80}, 094012 (2009)  [arXiv:0906.5333 [nucl-th]].
  %%CITATION = ARXIV:0906.5333;%%  %28 citations counted in INSPIRE as of 04 May 2013

%\cite{vanBeveren:2009fb}
\bibitem{vanBeveren:2009fb}
  E.~van Beveren and G.~Rupp,
  %``The X(4260) and possible confirmation of psi(3D), psi(5S), psi(4D), psi(6S) and psi(5D) in J/psi pi+ pi-,''
  arXiv:0904.4351 [hep-ph].
  %%CITATION = ARXIV:0904.4351;%%  %21 citations counted in INSPIRE as of 04 May 2013

%\cite{vanBeveren:2009jk}
\bibitem{vanBeveren:2009jk}
  E.~van Beveren and G.~Rupp,
  %``Interference effects in the X(4260) signal,''
  Phys.\ Rev.\ D {\bf 79}, 111501 (2009)  [arXiv:0905.1595 [hep-ph]].
  %%CITATION = ARXIV:0905.1595;%%  %17 citations counted in INSPIRE as of 04 May 2013

%\cite{vanBeveren:2010mg}
\bibitem{vanBeveren:2010mg}
  E.~van Beveren, G.~Rupp and J.~Segovia,
  %``A Very broad X(4260) and the resonance parameters of the \psi3D vector charmonium state,''
  Phys.\ Rev.\ Lett.\  {\bf 105} (2010) 102001
  [arXiv:1005.1010 [hep-ph]].
  %%CITATION = ARXIV:1005.1010;%%
  %11 citations counted in INSPIRE as of 08 Jul 2013

%\cite{Chen:2010nv}
\bibitem{Chen:2010nv}
  D.~-Y.~Chen, J.~He and X.~Liu,
  %``Nonresonant explanation for the Y(4260) structure observed in the $e^+e^-\to J/\psi\pi^+\pi^-$ process,''
  Phys.\ Rev.\ D {\bf 83} (2011) 054021
  [arXiv:1012.5362 [hep-ph]].
  %%CITATION = ARXIV:1012.5362;%%
  %5 citations counted in INSPIRE as of 08 Jul 2013

%\cite{Ebert:2005nc}
\bibitem{Ebert:2005nc}
  D.~Ebert, R.~N.~Faustov and V.~O.~Galkin,
  %``Masses of heavy tetraquarks in the relativistic quark model,''
  Phys.\ Lett.\ B {\bf 634}, 214 (2006)  [hep-ph/0512230].
  %%CITATION = HEP-PH/0512230;%%  %88 citations counted in INSPIRE as of 30 Apr 2013

%\cite{Zhang:2010mw}
\bibitem{Zhang:2010mw}
  J.~-R.~Zhang and M.~-Q.~Huang,
  %``The $P$-wave $[cs][\bar{c}\bar{s}]$ tetraquark state: $Y(4260)$ or $Y(4660)$?,''
  Phys.\ Rev.\ D {\bf 83}, 036005 (2011)  [arXiv:1011.2818 [hep-ph]].
  %%CITATION = ARXIV:1011.2818;%%  %4 citations counted in INSPIRE as of 11 Jul 2013

%\cite{Bali:2003tp}
\bibitem{Bali:2003tp}
  G.~S.~Bali,
  %``Lattice calculations of hadron properties,''
  Eur.\ Phys.\ J.\ A {\bf 19}, 1 (2004)  [hep-lat/0308015].
  %%CITATION = HEP-LAT/0308015;%%  %26 citations counted in INSPIRE as of 30 Apr 2013

%\cite{Wang:2007ea}
\bibitem{Wang:2007ea}
  X.~L.~Wang {\it et al.}  [Belle Collaboration],
  %``Observation of Two Resonant Structures in e+e- to pi+ pi- psi(2S) via Initial State Radiation at Belle,''
  Phys.\ Rev.\ Lett.\  {\bf 99}, 142002 (2007)  [arXiv:0707.3699 [hep-ex]].
  %%CITATION = ARXIV:0707.3699;%%  %201 citations counted in INSPIRE as of 30 Apr 2013

%\cite{Liu:2013vfa}
\bibitem{Liu:2013vfa}
  X.~-H.~Liu and G.~Li,
  %``Exploring the threshold behavior and implications on the nature of $Y(4260)$ and $Z_c(3900)$,''
  arXiv:1306.1384 [hep-ph].
  %%CITATION = ARXIV:1306.1384;%%

%\cite{Li:2013yka}
\bibitem{Li:2013yka}
  X.~Li and M.~B.~Voloshin,
  %``Suppression of the S-wave production of (3/2)^+ + (1/2)^- heavy meson pairs in e^+e^- annihilation,''
  Phys.\ Rev.\  D {\bf 588}, 034012 (2013). arXiv:1307.1072 [hep-ph].
  %%CITATION = ARXIV:1307.1072;%%

%\cite{Li:1996yn}
\bibitem{Li:1996yn}
  X.~Q.~Li, D.~V.~Bugg and B.~S.~Zou,
  %``A Possible explanation of the 'rho pi puzzle' in J / psi, psi-prime
  %decays,''
  Phys.\ Rev.\  D {\bf 55}, 1421 (1997).
  %%CITATION = PHRVA,D55,1421;%%

%\cite{Zhao:2006dv}
\bibitem{Zhao:2006dv}
  Q.~Zhao and B.~S.~Zou,
  %``On the near-threshold enhancement in J/ psi ---> gamma X with X ---> omega
  %phi,''
  Phys.\ Rev.\  D {\bf 74}, 114025 (2006)
  [arXiv:hep-ph/0606196].
  %%CITATION = PHRVA,D74,114025;%%

%\cite{Zhao:2006cx}
\bibitem{Zhao:2006cx}
  Q.~Zhao,
  %``A Study of the Okubo-Zweig-Iizuka rule violations in eta(c) ---> VV,''
  Phys.\ Lett.\  B {\bf 636}, 197 (2006)
  [arXiv:hep-ph/0602216].
  %%CITATION = PHLTA,B636,197;%%

%\cite{Li:2011ssa}
\bibitem{Li:2011ssa}
  G.~Li and Q.~Zhao,
  %``Revisit the radiative decays of $J/\psi$ and $\psi'\to \gamma\eta_c (\gamma\eta_c^\prime)$,''
  Phys.\ Rev.\ D {\bf 84}, 074005 (2011)  [arXiv:1107.2037 [hep-ph]].
  %%CITATION = ARXIV:1107.2037;%%  %12 citations counted in INSPIRE as of 13 Jun 2013

%\cite{Li:2007ky}
\bibitem{Li:2007ky}
  G.~Li, Q.~Zhao and C.~-H.~Chang,
  %``Decays of J/ psi and psi-prime into vector and pseudoscalar meson and the pseudoscalar glueball-q anti-q mixing,''
  J.\ Phys.\ G {\bf 35}, 055002 (2008)  [hep-ph/0701020].
  %%CITATION = HEP-PH/0701020;%%  %34 citations counted in INSPIRE as of 13 Jun 2013

%\cite{Wang:2012mf}
\bibitem{Wang:2012mf}
  Q.~Wang, G.~Li and Q.~Zhao,
  %``Open charm effects in the explanation of the long-standing '$\rho\pi$ puzzle',''
  Phys.\ Rev.\ D {\bf 85}, 074015 (2012)  [arXiv:1201.1681 [hep-ph]].
  %%CITATION = ARXIV:1201.1681;%%  %6 citations counted in INSPIRE as of 13 Jun 2013

%\cite{Li:2012as}
\bibitem{Li:2012as}
  G.~Li, F.~L.~Shao, C.~W.~Zhao and Q.~Zhao,
  %``$Z_b/Z_b^\prime \to \Upsilon\pi$ and $h_b \pi$ decays in intermediate meson loops model,''
  Phys.\ Rev.\ D {\bf 87}, 034020 (2013) arXiv:1212.3784 [hep-ph].
  %%CITATION = ARXIV:1212.3784;%%  %3 citations counted in INSPIRE as of 13 Jun 2013

%\cite{Li:2007xr}
\bibitem{Li:2007xr}
  G.~Li and Q.~Zhao,
  %``Hadronic loop contributions to J / psi and psi-prime radiative decays into gamma eta(c) or gamma eta(c)-prime,''
  Phys.\ Lett.\ B {\bf 670}, 55 (2008)  [arXiv:0709.4639 [hep-ph]].
  %%CITATION = ARXIV:0709.4639;%%  %27 citations counted in INSPIRE as of 13 Jun 2013

%\cite{Achasov:1990gt}
\bibitem{Achasov:1990gt}
  N.~N.~Achasov and A.~A.~Kozhevnikov,
  %``Direct decays of heavy quarkonia,''
  Phys.\ Lett.\ B {\bf 260}, 425 (1991).
  %%CITATION = PHLTA,B260,425;%%
  %5 citations counted in INSPIRE as of 06 Mar 2013
%\cite{Achasov:1991qp}

\bibitem{Achasov:1991qp}
  N.~N.~Achasov and A.~A.~Kozhevnikov,
  %``Decays of heavy quarkonia which violate the OZI rule,''
  JETP Lett.\  {\bf 54}, 193 (1991)
  [Pisma Zh.\ Eksp.\ Teor.\ Fiz.\  {\bf 54}, 197 (1991)].
  %%CITATION = JTPLA,54,193;%%
  %3 citations counted in INSPIRE as of 06 Mar 2013
%\cite{Achasov:1994vh}

\bibitem{Achasov:1994vh}
  N.~N.~Achasov and A.~A.~Kozhevnikov,
  %``Dynamical violation of the OZI rule and G parity in the decays of heavy quarkonia,''
  Phys.\ Rev.\ D {\bf 49}, 275 (1994).
  %%CITATION = PHRVA,D49,275;%%
  %16 citations counted in INSPIRE as of 06 Mar 2013

%\cite{Achasov:2005qb}
\bibitem{Achasov:2005qb}
  N.~N.~Achasov and A.~A.~Kozhevnikov,
  %``Branching ratios for the decays of $\psi$(3770) and
  %$\Upsilon$(10580) mesons to a pair of light hadrons,''
  Phys.\ Atom.\ Nucl.\  {\bf 69}, 988 (2006).
%  [Yad.\ Fiz.\  {\bf 69}, 1017 (2006)]
  [arXiv:hep-ph/0505146].
  %%CITATION = YAFIA,69,1017;%%

%\cite{Zhang:2009kr}
\bibitem{Zhang:2009kr}
  Y.~J.~Zhang, G.~Li and Q.~Zhao,
  %``Towards a dynamical understanding of the non-D anti-D decay of psi(3770),''
  Phys.\ Rev.\ Lett.\  {\bf 102}, 172001 (2009)
  [arXiv:0902.1300 [hep-ph]].
  %%CITATION = PRLTA,102,172001;%%

%\cite{Liu:2009dr}
\bibitem{Liu:2009dr}
  X.~Liu, B.~Zhang and X.~Q.~Li,
  %``The puzzle of excessive non-$D\bar D$ component of the inclusive
  %$\psi(3770)$ decay and the long-distant contribution,''
  Phys.\ Lett.\  B {\bf 675}, 441 (2009)
  [arXiv:0902.0480 [hep-ph]].
  %%CITATION = PHLTA,B675,441;%%

%\cite{Wu:2007jh}
\bibitem{Wu:2007jh}
  J.~J.~Wu, Q.~Zhao and B.~S.~Zou,
  %``Possibility of measuring a0(980)-f0(980) mixing from J/psi ---> phi
  %a0(980),''
  Phys.\ Rev.\  D {\bf 75}, 114012 (2007)
  [arXiv:0704.3652 [hep-ph]].
  %%CITATION = PHRVA,D75,114012;%%

%\cite{Liu:2006dq}
\bibitem{Liu:2006dq}
  X.~Liu, X.~Q.~Zeng and X.~Q.~Li,
  %``Study on contributions of hadronic loops to decays of J/psi ---> vector +
  %pseudoscalar mesons,''
  Phys.\ Rev.\  D {\bf 74}, 074003 (2006)
  [arXiv:hep-ph/0606191].
  %%CITATION = PHRVA,D74,074003;%%

%\cite{Cheng:2004ru}
\bibitem{Cheng:2004ru}
  H.~Y.~Cheng, C.~K.~Chua and A.~Soni,
  %``Final state interactions in hadronic B decays,''
  Phys.\ Rev.\  D {\bf 71}, 014030 (2005)
  [arXiv:hep-ph/0409317].
  %%CITATION = PHRVA,D71,014030;%%

%\cite{Anisovich:1995zu}
\bibitem{Anisovich:1995zu}
  V.~V.~Anisovich, D.~V.~Bugg, A.~V.~Sarantsev and B.~S.~Zou,
  %``Upsilon (3S) ---> Upsilon (1S) pi pi decay: Is the pi pi spectrum puzzle an
  %indication of a b anti-b q anti-q resonance?,''
  Phys.\ Rev.\  D {\bf 51}, 4619 (1995).
  %%CITATION = PHRVA,D51,4619;%%

%\cite{Zhao:2005ip}
\bibitem{Zhao:2005ip}
  Q.~Zhao, B.~s.~Zou and Z.~b.~Ma,
  %``Glueball-Q anti-Q mixing and Okuba-Zweig-Iizuka rule violation in the
  %hadronic decays of heavy quarkonia,''
  Phys.\ Lett.\  B {\bf 631}, 22 (2005)
  [arXiv:hep-ph/0508088].
  %%CITATION = PHLTA,B631,22;%%

%\cite{Li:2007au}
\bibitem{Li:2007au}
  G.~Li, Q.~Zhao and B.~S.~Zou,
  %``Isospin violation in phi, J/psi, psi-prime ---> omega pi0 via hadronic
  %loops,''
  Phys.\ Rev.\  D {\bf 77}, 014010 (2008)
  [arXiv:0706.0384 [hep-ph]].
  %%CITATION = PHRVA,D77,014010;%%

%\cite{Liu:2009vv}
\bibitem{Liu:2009vv}
  X.~H.~Liu and Q.~Zhao,
  %``The Evasion of helicity selection rule in chi(c1) ---> VV and chi(c2) --->
  %VP via intermediate charmed meson loops,''
  Phys.\ Rev.\  D {\bf 81}, 014017 (2010)
  [arXiv:0912.1508 [hep-ph]].
  %%CITATION = PHRVA,D81,014017;%%

%\cite{Wang:2012wj}
\bibitem{Wang:2012wj}
  Q.~Wang, X.~-H.~Liu and Q.~Zhao,
  %``Updated Study of the $\eta_c$ and $\eta_c^\Prime$ Decays into Light Vector Mesons,''
  Phys.\ Lett.\ B {\bf 711}, 364 (2012)  [arXiv:1202.3026 [hep-ph]].
  %%CITATION = ARXIV:1202.3026;%%

%\cite{Liu:2010um}
\bibitem{Liu:2010um}
  X.~H.~Liu and Q.~Zhao,
  %``Further study of the helicity selection rule evading mechanism in $\eta_c$,
  %$\chi_{c0}$ and $h_c$ decaying to baryon anti-baryon pairs,''
  J.\ Phys.\ G {\bf 38}, 035007 (2011)
  [arXiv:1004.0496 [hep-ph]].
  %%CITATION = JPHGB,G38,035007;%%

%\cite{Guo:2009wr}
\bibitem{Guo:2009wr}
  F.~-K.~Guo, C.~Hanhart and U.~-G.~Mei{\ss}ner,
  %``On the extraction of the light quark mass ratio from the decays psi-prime ---> J/psi pi0 (eta),''
  Phys.\ Rev.\ Lett.\  {\bf 103}, 082003 (2009)  [Erratum-ibid.\  {\bf 104}, 109901 (2010)]  [arXiv:0907.0521 [hep-ph]].
  %%CITATION = ARXIV:0907.0521;%%  %23 citations counted in INSPIRE as of 06 May 2013

%\cite{Guo:2010zk}
\bibitem{Guo:2010zk}
  F.~K.~Guo, C.~Hanhart, G.~Li, U.~G.~Mei{\ss}ner and Q.~Zhao,
  %``Novel analysis of the decays psi' -> h_c pi^0 and eta_c'-> chi_{c0} pi^0,''
  Phys.\ Rev.\  D {\bf 82}, 034025 (2010)
  [arXiv:1002.2712 [hep-ph]].
  %%CITATION = PHRVA,D82,034025;%%

%\cite{Guo:2010ak}
\bibitem{Guo:2010ak}
  F.~K.~Guo, C.~Hanhart, G.~Li, U.~G.~Mei{\ss}ner and Q.~Zhao,
  %``Effect of charmed meson loops on charmonium transitions,''
  Phys.\ Rev.\  D {\bf 83}, 034013 (2011)
  [arXiv:1008.3632 [hep-ph]].
  %%CITATION = PHRVA,D83,034013;%%

%\cite{Li:2013xia}
\bibitem{Li:2013xia}
  G.~Li,
  %``Hidden charmonium decays of $Z_c$ in intermediate meson loops model,''
  arXiv:1304.4458 [hep-ph].  %%CITATION = ARXIV:1304.4458;
  %%  %1 citations counted in INSPIRE as of 30 Apr 2013

%\cite{Brambilla:2004wf}
\bibitem{Brambilla:2004wf}
  N.~Brambilla {\it et al.}  [Quarkonium Working Group Collaboration],
  %``Heavy quarkonium physics,''
  hep-ph/0412158.
  %%CITATION = HEP-PH/0412158;%%  %582 citations counted in INSPIRE as of 12 Apr 2013

%\cite{Brambilla:2004jw}
\bibitem{Brambilla:2004jw}
  N.~Brambilla, A.~Pineda, J.~Soto and A.~Vairo,
  %``Effective field theories for heavy quarkonium,''
  Rev.\ Mod.\ Phys.\  {\bf 77}, 1423 (2005)  [hep-ph/0410047].
  %%CITATION = HEP-PH/0410047;%%  %230 citations counted in INSPIRE as of 12 Apr 2013

%\cite{Colangelo:2003sa}
\bibitem{Colangelo:2003sa}
  P.~Colangelo, F.~De Fazio and T.~N.~Pham,
  %``Nonfactorizable contributions in B decays to charmonium: The case of B-
  %--> K- h/c,''
  Phys.\ Rev.\  D {\bf 69}, 054023 (2004)
  [arXiv:hep-ph/0310084].
  %%CITATION = PHRVA,D69,054023;%%

%\cite{Casalbuoni:1996pg}
\bibitem{Casalbuoni:1996pg}
  R.~Casalbuoni, A.~Deandrea, N.~Di Bartolomeo, R.~Gatto, F.~Feruglio and G.~Nardulli,
  %``Phenomenology of heavy meson chiral Lagrangians,''
  Phys.\ Rept.\  {\bf 281}, 145 (1997)
  [arXiv:hep-ph/9605342].
  %%CITATION = PRPLC,281,145;%%

%\cite{Colangelo:2002mj}
\bibitem{Colangelo:2002mj}
  P.~Colangelo, F.~De Fazio and T.~N.~Pham,
  %``B- ---> K- (chi(c0)) decay from charmed meson rescattering,''
  Phys.\ Lett.\ B {\bf 542}, 71 (2002)  [hep-ph/0207061].
  %%CITATION = HEP-PH/0207061;%%

%\cite{Veliev:2010gb}
\bibitem{Veliev:2010gb}
  E.~V.~Veliev, H.~Sundu, K.~Azizi and M.~Bayar,
  %``Scalar Quarkonia at Finite Temperature,''
  Phys.\ Rev.\ D {\bf 82}, 056012 (2010)  [arXiv:1003.0119 [hep-ph]].
  %%CITATION = ARXIV:1003.0119;%%

%\cite{Casalbuoni:1992gi}
\bibitem{Casalbuoni:1992gi}
  R.~Casalbuoni, A.~Deandrea, N.~Di Bartolomeo, R.~Gatto, F.~Feruglio and G.~Nardulli,
  %``Light vector resonances in the effective chiral Lagrangian for heavy mesons,''  
  Phys.\ Lett.\ B {\bf 292}, 371 (1992)  [hep-ph/9209248].  
  %%CITATION = HEP-PH/9209248;%%  %89 citations counted in INSPIRE as of 04 May 2013

%\cite{Casalbuoni:1992dx}
\bibitem{Casalbuoni:1992dx}
  R.~Casalbuoni, A.~Deandrea, N.~Di Bartolomeo, R.~Gatto, F.~Feruglio and G.~Nardulli,
  %``Effective Lagrangian for heavy and light mesons: Semileptonic decays,''  
  Phys.\ Lett.\ B {\bf 299}, 139 (1993)  [hep-ph/9211248].  
  %%CITATION = HEP-PH/9211248;%%  %154 citations counted in INSPIRE as of 04 May 2013

%\cite{Isola:2003fh}
\bibitem{Isola:2003fh}
  C.~Isola, M.~Ladisa, G.~Nardulli and P.~Santorelli,
  %``Charming penguins in B ---> K* pi, K(rho, omega, phi) decays,''
  Phys.\ Rev.\ D {\bf 68}, 114001 (2003)  [hep-ph/0307367].
  %%CITATION = HEP-PH/0307367;%%  %66 citations counted in INSPIRE as of 04 May 2013

%\cite{Burdman:1992gh}
\bibitem{Burdman:1992gh}
  G.~Burdman and J.~F.~Donoghue,
  %``Union of chiral and heavy quark symmetries,''
  Phys.\ Lett.\ B {\bf 280}, 287 (1992).
  %%CITATION = PHLTA,B280,287;%%  %442 citations counted in INSPIRE as of 04 May 2013

%\cite{Yan:1992gz}
\bibitem{Yan:1992gz}
  T.~-M.~Yan, H.~-Y.~Cheng, C.~-Y.~Cheung, G.~-L.~Lin, Y.~C.~Lin and H.~-L.~Yu,
  %``Heavy quark symmetry and chiral dynamics,''
  Phys.\ Rev.\ D {\bf 46}, 1148 (1992)  [Erratum-ibid.\ D {\bf 55}, 5851 (1997)].
  %%CITATION = PHRVA,D46,1148;%%  %445 citations counted in INSPIRE as of 04 May 2013

%\cite{Falk:1992cx}
\bibitem{Falk:1992cx}
  A.~F.~Falk and M.~E.~Luke,
  %``Strong decays of excited heavy mesons in chiral perturbation theory,''
  Phys.\ Lett.\ B {\bf 292}, 119 (1992)  [hep-ph/9206241].
  %%CITATION = HEP-PH/9206241;%%  %145 citations counted in INSPIRE as of 04 May 2013

%\cite{Deandrea:1999pa}
\bibitem{Deandrea:1999pa}
  A.~Deandrea, R.~Gatto, G.~Nardulli and A.~D.~Polosa,
  %``Couplings of pions to higher positive parity heavy mesons,''
  JHEP {\bf 9902}, 021 (1999)  [hep-ph/9901266].
  %%CITATION = HEP-PH/9901266;%%  %15 citations counted in INSPIRE as of 04 May 2013

%\cite{Weinberg:1965zz}
\bibitem{Weinberg:1965zz}
  S.~Weinberg,
  %``Evidence That the Deuteron Is Not an Elementary Particle,''
  Phys.\ Rev.\  {\bf 137} (1965) B672.

\bibitem{Baru:2003qq}
  V.~Baru  {\it et al.},
  %``Evidence that the a(0)(980) and f(0)(980) are not elementary particles,''
  Phys.\ Lett.\ B {\bf 586}, 53 (2004).
%  [hep-ph/0308129].

%\cite{Locher:1993cc}
\bibitem{Locher:1993cc}
  M.~P.~Locher, Y.~Lu and B.~S.~Zou,
  %``Rates for the reactions anti-p p ---> pi phi and gamma phi,''
  Z.\ Phys.\ A {\bf 347}, 281 (1994)  [nucl-th/9311021].
  %%CITATION = NUCL-TH/9311021;%%  %59 citations counted in INSPIRE as of 14 Jun 2013

%\cite{Li:1996cj}
\bibitem{Li:1996cj}
  X.~-Q.~Li and B.~-S.~Zou,
  %``Significance of single pion exchange inelastic final state interaction for D ---> V P processes,''
  Phys.\ Lett.\ B {\bf 399}, 297 (1997)  [hep-ph/9611223].
  %%CITATION = HEP-PH/9611223;%%  %20 citations counted in INSPIRE as of 14 Jun 2013

%\cite{Cleven:2011gp}
\bibitem{Cleven:2011gp}
  M.~Cleven, F.~-K.~Guo, C.~Hanhart and U.~-G.~Mei{\ss}ner,
  %``Bound state nature of the exotic $Z_b$ states,''
  Eur.\ Phys.\ J.\ A {\bf 47}, 120 (2011)  [arXiv:1107.0254 [hep-ph]].
  %%CITATION = ARXIV:1107.0254;%%  %33 citations counted in INSPIRE as of 23 Aug 2013

%\cite{Bai:2001ct}
\bibitem{Bai:2001ct}
  J.~Z.~Bai {\it et al.}  [BES Collaboration],
  %``Measurements of the cross-section for e+ e ---> hadrons at center-of-mass energies from 2-GeV to 5-GeV,''
  Phys.\ Rev.\ Lett.\  {\bf 88}, 101802 (2002)  [hep-ex/0102003].
  %%CITATION = HEP-EX/0102003;%%  %250 citations counted in INSPIRE as of 28 Aug 2013

%\cite{Mo:2006ss}
\bibitem{Mo:2006ss}
  X.~H.~Mo, G.~Li, C.~Z.~Yuan, K.~L.~He, H.~M.~Hu, J.~H.~Hu, P.~Wang and Z.~Y.~Wang,
  %``Determining the upper limit of Gamma(ee) for the Y(4260),''
  Phys.\ Lett.\ B {\bf 640}, 182 (2006)  [hep-ex/0603024].
  %%CITATION = HEP-EX/0603024;%%  %39 citations counted in INSPIRE as of 28 Aug 2013

\bibitem{Yuan:2013}
Changzheng Yuan, Talks in the XXVI International Symposium on Lepton
Photon Interactions at High Energies, San Francisco, USA, 2013

%\cite{Pakhlova:2010ek}
\bibitem{Pakhlova:2010ek}
  G.~Pakhlova {\it et al.}  [Belle Collaboration],
  %``Measurement of $e^+e^-\to D_s^{(*)+} D_s^{(*)-}$ cross sections near threshold using initial-state radiation,''
  Phys.\ Rev.\ D {\bf 83}, 011101 (2011)  [arXiv:1011.4397 [hep-ex]].
  %%CITATION = ARXIV:1011.4397;%%  %7 citations counted in INSPIRE as of 28 Aug 2013

%\cite{Pakhlova:2008vn}
\bibitem{Pakhlova:2008vn}
  G.~Pakhlova {\it et al.}  [Belle Collaboration],
  %``Observation of a near-threshold enhancement in the e+e- ---> Lambda+(c) Lambda-(c) cross section using initial-state radiation,''
  Phys.\ Rev.\ Lett.\  {\bf 101}, 172001 (2008)  [arXiv:0807.4458 [hep-ex]].
  %%CITATION = ARXIV:0807.4458;%%  %81 citations counted in INSPIRE as of 28 Aug 2013

%\cite{Pakhlova:2009jv}
\bibitem{Pakhlova:2009jv}
  G.~Pakhlova {\it et al.}  [Belle Collaboration],
  %``Measurement of the e+ e- ---> D0 D*- pi+ cross section using initial-state radiation,''
  Phys.\ Rev.\ D {\bf 80}, 091101 (2009)
  [arXiv:0908.0231 [hep-ex]].
  %%CITATION = ARXIV:0908.0231;%%
  %35 citations counted in INSPIRE as of 23 Aug 2013

%\cite{Pakhlova:2007fq}
\bibitem{Pakhlova:2007fq}
  G.~Pakhlova {\it et al.}  [Belle Collaboration],
  %``Observation of psi (4415) ---> D anti-D*(2)(2460) decay using initial-state radiation,''
  Phys.\ Rev.\ Lett.\  {\bf 100}, 062001 (2008)
  [arXiv:0708.3313 [hep-ex]].
  %%CITATION = ARXIV:0708.3313;%%
  %55 citations counted in INSPIRE as of 23 Aug 2013

%\cite{Abe:2006fj}
\bibitem{Abe:2006fj}
  K.~Abe {\it et al.}  [Belle Collaboration],
  %``Measurement of the near-threshold e+ e- ---> D(*)+- D(*)-+ cross section using initial-state radiation,''
  Phys.\ Rev.\ Lett.\  {\bf 98}, 092001 (2007)
  [hep-ex/0608018].
  %%CITATION = HEP-EX/0608018;%%
  %102 citations counted in INSPIRE as of 23 Aug 2013

%\cite{Pakhlova:2008zza}
\bibitem{Pakhlova:2008zza}
  G.~Pakhlova {\it et al.}  [Belle Collaboration],
  %``Measurement of the near-threshold e+ e- ---> D anti-D cross section using initial-state radiation,''
  Phys.\ Rev.\ D {\bf 77}, 011103 (2008)
  [arXiv:0708.0082 [hep-ex]].
  %%CITATION = ARXIV:0708.0082;%%
  %84 citations counted in INSPIRE as of 23 Aug 2013

%\cite{Aubert:2006mi}
\bibitem{Aubert:2006mi}
  B.~Aubert {\it et al.}  [BaBar Collaboration],
  %``Study of the Exclusive Initial-State Radiation Production of the D anti-D System,''
  Phys.\ Rev.\ D {\bf 76}, 111105 (2007)
  [hep-ex/0607083].
  %%CITATION = HEP-EX/0607083;%%
  %69 citations counted in INSPIRE as of 23 Aug 2013

%\cite{Aubert:2009aq}
\bibitem{Aubert:2009aq}
  B.~Aubert {\it et al.}  [BaBar Collaboration],
  %``Exclusive Initial-State-Radiation Production of the D anti-D, D* anti-D*, and D* anti-D* Systems,''
  Phys.\ Rev.\ D {\bf 79}, 092001 (2009)
  [arXiv:0903.1597 [hep-ex]].
  %%CITATION = ARXIV:0903.1597;%%
  %64 citations counted in INSPIRE as of 23 Aug 2013

%\cite{Guo:2012tg}
\bibitem{Guo:2012tg}
  F.~-K.~Guo and U.~-G.~Mei{\ss}ner,
  %``Light quark mass dependence in heavy quarkonium physics,''
  Phys.\ Rev.\ Lett.\  {\bf 109}, 062001 (2012)  [arXiv:1203.1116 [hep-ph]].
  %%CITATION = ARXIV:1203.1116;%%  %4 citations counted in INSPIRE as of 06 Jul 2013

%\cite{Dashen:1969eg}
\bibitem{Dashen:1969eg}
  R.~F.~Dashen,
  %``Chiral SU(3) x SU(3) as a symmetry of the strong interactions,''
  Phys.\ Rev.\  {\bf 183}, 1245 (1969).
  %%CITATION = PHRVA,183,1245;%%  %571 citations counted in INSPIRE as of 05 May 2013







%%%%%%%%%%%%%%%%%%%%%%%%%%%%%%%%%%%%%%%%%%%%%%%%%%%%%%%%%%%%%
%%%%%%%%%%%%%%%%%%%%%%%%%%%%%%%%%%%%%%%%%%%%%%%%%%%%%%%%%%%%%





\end{thebibliography}
\end{document}